\documentclass[a4paper]{article}
\usepackage[T1]{fontenc}
\usepackage[utf8]{inputenc}
\usepackage[italian,english]{babel}
\usepackage{amsmath}
\usepackage{graphicx}
\usepackage{anysize}
\usepackage{float}
\usepackage{amsmath}
\usepackage{amssymb}
\usepackage{amsfonts}
\usepackage{bm}
\usepackage{array}
\usepackage{graphicx}
\usepackage{epsfig}
\usepackage{multirow}
\usepackage{multicol}
\usepackage{longtable}
\usepackage{ifthen}
\usepackage{lscape}
\usepackage{alphalph}
\usepackage{attachfile2}
\usepackage {textcomp}
\graphicspath{{figs/}}
\normalmarginpar
\begin{document}

  \begin{titlepage}
  \begin{center}

\textbf{On super-elastic collisions between magnetized plasmoids in the heliosphere}
\

A. Di Vita \footnote{Università di Genova, Via Montallegro 1, 16145 Genova, Italy - 2017, April the 20th}, 

  \end{center}

\begin{abstract}
Recently, a unique collision between two large-scale magnetized plasmoids produced by coronal mass ejections in the heliosphere has been observed [C. Shen et al., Nature Physics \textbf{8}, 923–928 (2012)]. Results suggest that the collision is super-elastic, i.e. the total linear kinetic energy of the two plasmoids after the collision is larger than before the collision, and that an anti-correlation exists, i.e. the lower the initial relative velocity of the plasmoids, the larger the relative increase in total kinetic energy. Following an old suggestion of [W. H. Bostick, IEEE Trans. Plasma Science \textbf{PS-14} 703-717 (1986)], here we start from first principles, retrieve some results of [S. Ohsaki et al.,  Ap. J. Lett. \textbf{559} L61 (2001)] and [D. Kagan et al., Mon. Not. R. Astron. Soc. \textbf{406} 1140-1145 (2010)] and show that the anti-correlation is just a consequence of the properties of Joule and viscous dissipation inside the plasmoids. On the other end, if the initial relative velocity of the plasmoids is greater than the Alfvèn velocity times the global reconnection rate, then the plasmoids merge.
\end{abstract}


\end{titlepage}
 
\section{The problem}
\label{SEC1}
 
Coronal mass ejections (CMEs) give birth to large-scale plasmoids, originating from the solar atmosphere and expanding and propagating into the heliosphere \cite{ShenNature} \cite{Cargill}. The occurrence rate of CMEs is about 4-5 CMEs per day at solar maximum \cite{Mishra}, so that encounters and interactions between plasmoids are unavoidable. Nevertheless, inter-plasmoid collisions are far from understood. While collisions between blobs of ordinary gases are heavily affected by mixing, cross-field diffusion is effectively prohibited by sufficiently strong magnetic field. 

Experiments show that initially well-distinct plasmoids preserve their identity after a collision; historically, these experiments led to the very definition of ‘plasmoid’ as a ‘plasma magnetic entity’ with well-distinguished identity and geometrical structure \cite{Bostick}. In the lab, conventional magnetohydrodynamics fails to provide adequate macroscopic description of isolated plasmoids \cite{Di Vita00} – let alone their mutual interaction. In space, such description requires detailed knowledge of energy balance and state equation \cite{Cargill}. 

Recent observation of a collision between plasmoids \cite{ShenNature} suggests that such collisions may be super-elastic, i.e. the total \textit{linear} kinetic energy of the colliding plasmoids after the collision is larger than before the collision. A similar phenomenon is observed in collisions of spheres with elastoplastic plates, where rotational kinetic energy may be transferred into linear kinetic energy \cite{Louge}. This analogy suggests that the linear kinetic energy in collisions between plasmoids may increase at the expense of the energy of some other degree of freedom, and magnetic energy is an obvious candidate. A preliminary investigation of energy balance supports this point of view; moreover, it has been suggested that the lower the impact velocity, the larger the relative increase of the linear kinetic energy \cite{ShenNature}; in contrast, if the relative velocity of colliding plasmoids is too large, then no elastic scattering occurs \cite{TURNON}, and the plasmoids may merge into each other. 

The aim of this work is to investigate this suggestion. We invoke the analogy \cite{Bostick1} between plasmoids in the lab and space plasmoids and show how recent progress \cite{Di Vita00} \cite{Di Vita01} in the analytical description of the former provide information on the latter. In particular, we show that quantities usually invoked in Hall MHD \cite{Witalis} lead to a simple macroscopic description of interacting plasmoids. Some relevant properties of our plasmoids are discussed in Sec. 2. Secs. 3 and 4 discuss the structure of single plasmoids. Sec. 5 describes plasmoid-plasmoid scattering (heneceforth referred to just as 'scattering'). Sec. 6 discusses the possible merging of plasmoids. Conclusions are drawn in Sec. 7. Unless otherwise specified, SI units are used for all quantities but temperatures (which are in eV).

\section{Some facts about our plasmoids}
\label{SEC2}

At 1 A.U., typical values of the absolute value $ \vert \textbf{B} \vert $ of the magnetic field $ \textbf{B} $, the electron density $ n_{e} $, the electron temperature $ T_{e} $ and the typical center-of-mass speed $ v_{0} $ in a plasmoid like those described in \cite{ShenNature} (whose scattering takes about $ \Delta t \approx 16 $ hours) are $ B_{\left(1 A.U.\right)} \approx 10^{-8} $ T, $ n_{e \left(1 A.U.\right)} \approx 5 \cdot 10^{6} $ m$ ^{-3} $, $ T_{e \left(1 A.U.\right)} \approx 10 $ eV and $ v_{0 \left(1 A.U.\right)} \approx 5 \cdot 10^{5} $ m/s respectively. These data are not in contradiction with the observations e.g. of \cite{Jian}. According to the data displayed in Fig. 4 of \cite{ShenNature}, the typical value $ L_{\left(1 A.U.\right)} $ of the plasmoid radius $ L $ is in the range 5 – 10 times the radius $ R_{S} $ of the Sun, i.e. between $ 3.5 \cdot 10^9 $ m and $ 7 \cdot 10^9 $ m.  We assume that the plasmoid is made of pure, fully ionized hydrogen, so that the ion density $ n_{i} $ is equal to $ n_{e} $. Correspondingly, reasonable estimates of the values of the ratio $ \beta $ of plasma pressure $ p $ and magnetic pressure, the ion Hall parameter $ \Lambda_{i} $, the Lundquist number $ S $ and the Hartmann number $ Ha $ (computed with the help of the  parallel viscosity coefficient $ \eta_{\left(V 0 \right)} $ and Spitzer’s parallel electric resistivity $ \eta_{//} $ respectively, see below) are $ \beta_{\left(1 A.U.\right)} \approx 0.4, \Lambda_{i \left(1 A.U.\right)} \approx 10^7, S_{\left(1 A.U.\right)} \approx 2 \cdot 10^{13}$ and $ Ha_{\left(1 A.U.\right)} \approx 10^{6} $ respectively. Following \cite{TURNON}, here and in the following we do not take into account formation of CME-related shock waves (with the exception of Appendix \ref{QUAL3}).

It is useful, for comparison, to look at what happens in the initial period of plasmoid life, i.e. when it is still located near the Sun. At a distance of 1.8 solar radii from the Sun center, reasonable values (and ranges) of $ \vert \textbf{B} \vert $, $ n_{e} $, $ T_{e} $,  $ v_{0} $, $ L $, $ \beta $, $ \Lambda_{i} $, $ S $ and $ Ha $ are $ B_{\left(init\right)} \approx 10^{-4} $ T, $ n_{e \left(init\right)} \approx 4 \cdot 10^{13} $ m$ ^{-3} $, $ T_{e \left(init\right)} \approx 30 $ eV, $ v_{0 \left(init\right)} \approx 2.4 \cdot 10^{5} - 4.1 \cdot 10^{5} $ m/s, $ L_{\left(init\right)} \approx 0.5 R_{S} - 0.6 R_{S}$, $ \beta_{\left(init\right)} \approx 0.06 $ (see p. 6 of \cite{ZHOU} and Table 1 of \cite{COULD}), $ \Lambda_{i \left(init\right)} \approx 5 \cdot 10^4$, $ S_{\left(init\right)} \approx 5 \cdot 10^{13} $ and $ Ha_{\left(init\right)} \approx 6 \cdot 10^{8} $ respectively. Table \ref{TABELLASINOTTICA} displays relevant data.

\begin{table}[h]  
\caption{Typical parameters for plasmoids near and far from the Sun}
\centering
\begin{tabular}{|c|c|c|} 
\hline 
$ \mbox{Distance from the Sun center} $ & $ 1.8 R_{S} $  & $ 1 A.U. $ \\ 
\hline 
$ B \left( T \right) $ & $ 10^{-4} $ &  $  10^{-8} $  \\ 
\hline 
$ n_{e} \left( m^{-3} \right)$ & $ 4 \cdot 10^{13}  $ & $ 5 \cdot 10^{6} $ \\ 
\hline 
$ T_{e} \left( eV \right)$ & $ 30 $ & $ 10 $ \\ 
\hline 
$ v_{0} \left( m \cdot s^{-1} \right)$ & $ 2.4 \cdot 10^{5} - 4.1 \cdot 10^{5} $ & $ 5 \cdot 10^{5} $ \\ 
\hline 
$ L \left( m \right) $ & $ 3.5 \cdot 10^8 - 4.2 \cdot 10^8 $ & $ 3.5 \cdot 10^9 - 7 \cdot 10^9 $ \\ 
\hline 
$ \beta $ & $ 0.06 $ & $ 0.4 $ \\ 
\hline 
$ \Lambda_{i} $ & $ 5 \cdot 10^4 $ & $ 10^7 $ \\ 
\hline 
$ S $ & $ 5 \cdot 10^{13} $ & $ 2 \cdot 10^{13} $ \\ 
\hline 
$ Ha $ & $ 6 \cdot 10^{8} $ & $ 10^{6} $ \\ 
\hline 
\end{tabular} 
\label{TABELLASINOTTICA}
\end{table}

Remarkably, both $ \Lambda_{i} $, $ S $ and $ Ha $ remain $ \gg 1 $ all along the path from the birth of the plasmoid up to the scattering with another plasmoid. Then, we may assume that dissipation leaves the essential features of the magnetic field topology unaffected during the flight from the Sun to the scattering. 

Moreover, near the Sun the magnetosonic speed $ c_{sA} $ is $ \approx $ hundreds of Km/s - see p. 6 of \cite{TURNON} - so that the typical time-scale $ \tau = \frac{L_{init}}{c_{sA}} \approx 10^3 $ s of transit of magnetosonic waves across the plasmoid is $ \ll $ the time-of-flight $ t_{TOF} \approx \frac{1 U.A.}{v_{0 \left(1 A.U.\right)}} \approx 10^6$ s elapsed between the birth of the plasmoid near the Sun and the scattering with another plasmoid, say 1 A.U. far from the Sun. This fact allows us to assume that – in the initial period of its life at least –  the evolution of the plasmoid is a succession of quasi-steady, relaxed states, the relaxation time $ \tau $ being $ \ll t_{TOF} $. Since the flight leaves the plasmoid structure unaffected, a description of the relaxed state of the plasmoid near the Sun may provide information concerning the initial conditions of the plasmoid motion during the scattering. The next Section describes such relaxed states. 

\section{The plasmoid as a relaxed state}
\label{SEC3}

\paragraph{Generalities.} In a plasmoid near the Sun, the fact that $ \beta_{\left(init\right)} \ll 1 $ makes $ \textbf{B} $ to be approximately force-free, i.e. $ \textbf{j} \wedge \textbf{B} \approx \nabla p \approx O \left( \beta \right) \approx 0 $, where $ \textbf{j} $ is the electric current density. Examples of such force-free configurations for $ \textbf{B} $ are the spheromak-like structure in eq. 8 of \cite{ZHOU} and the cylindrical solution of equations (1)-(3) of \cite{WANG}. Simulations show that the evolution of $ \textbf{B} $ is self-similar - see Fig. 7 of \cite{ZHOU} - and it is reasonable to assume the same for $ n_{e} $ and $ T $ - see e.g. the description of a non-isothermal plasmoid in Sec. II of \cite{Cargill}. For simplicity, we start with the assumption $ \nabla n_{e} = 0 $, take a common temperature $ T $ for ions and electrons and write $ \nabla T = 0 $ - or, equivalently, $ T \left( \textbf{x}, t \right) = T_{b} $ - everywhere across the plasmoid, where $ T_{b} $ is the value of $ T $ on the plasmoid boundary. (These oversimplifiying assumptions are to be dropped below). Given its spontaneous occurrence, relaxation is likely to involve some entropy-raising, dissipative phenomenon; if $ \nabla n_{e} = 0 $ and $ \nabla T = 0 $, then dissipation is ruled by viscous heating and Joule heating. We investigate these heating processes in order to obtain information about relaxation.

\paragraph{Viscous heating.} For the sake of simplicity we assume that turbulence (if any) leaves viscosity unaffected; this assumption is justified below. It turns out \cite{NRL} that ions rule viscous dissipation, as the ion mass $ m_{i} $ is $ \gg $ the electron mass $ m_{e} $; the same inequality ensures that the macroscopic velocity $ \textbf{v} $ (i.e. the velocity of the center-of-mass of a small mass element of the plasma) is mainly ruled by ions. The quantity $ \textbf{v} $ obeys the momentum balance:

\begin{equation} \label{NS}
\rho \dfrac{\partial \textbf{v} }{\partial t} 
+ \rho \left( \textbf{v} \cdot \nabla \right) \textbf{v}  
+ \nabla p 
+ \rho \nabla \phi_{g}
- \textbf{j} \wedge \textbf{B}
- \nabla \cdot \bm{\tau} = 0
\end{equation}

where $ t $, $ \rho $, $ \bm{\tau} $, $ \phi_{g} $ are the time, the mass density, the viscous stress tensor and the potential of the solar gravitational field respectively. (We dropped the dependence on $ \textbf{x} $ and $ t $ everywhere, for simplicity). The $ \left(i,j\right) $-th component $ \tau_{ij} $ of $ \bm{\tau} $ is $ \tau_{ij} = \eta_{\left( V \right) ijkl} \frac{\partial v_{i}}{\partial x_{j}}$ where $ \eta_{\left( V \right) ijkl} = \eta_{\left( V \right) jikl} = \eta_{\left( V \right) ijlk}$ ($ i, j, k, l = 1, 2, 3 $). In the $ \Lambda_{i} \gg 1 $ limit, the only relevant viscosity controls the variation in the direction of $ \textbf{B} $ of the component of $ \textbf{v} $ which is parallel to $ \textbf{B} $, and the tensor $ \bm{\eta}_{\left( V \right)} $ with components $ \eta_{\left( V \right) ijkl} $ reduces to $ 3 \eta_{\left(V 0 \right)} \left( \textbf{b} \textbf{b} - \frac{1}{3} \textbf{1} \right) \left( \textbf{b} \textbf{b} - \frac{1}{3} \textbf{1} \right)$. Here $ \textbf{b} \equiv \frac{\textbf{B}}{\vert \textbf{B} \vert}$, $ \textbf{1} $ is the identity tensor and $ \eta_{\left(V 0 \right)} $ is the parallel viscosity coefficient of unmagnetized plasmas, which does not depend on $ \Lambda_{i} $. The amount $ P_{v} $ of heat produced per unit volume and time by viscosity is: 

\begin{equation} \label{PV}
P_{v} = \tau_{ij} \dfrac{\partial v_{i}}{\partial x_{j}}
\end{equation}

where $ v_{i} $ and $ x_{j} $ are the i-th and the j-th component of $ \textbf{v} $ and of position $ \textbf{x} $ respectively. Finally, the large value of $ c_{sA} $ allows us to assume incompressibility, so that mass balance leads to:

\begin{equation} \label{incompressibility}
\nabla \cdot \textbf{v} = 0 \quad ; \quad \rho = \mbox{constant and uniform} 
\end{equation}

For future reference, here we recall the result of Sec. 344 of \cite{Lamb}: a steady state of an unmagnetized ($ \Lambda_{i} = Ha = 0 $), very viscous (Reynolds' number $ Re \ll 1 $) fluid which fills a fixed region $ \Omega $ of space and obeys both equations \eqref{NS} and \eqref{incompressibility} corresponds to a constrained minimum of the total viscous power $ \int_{\Omega} P_{v} d^3\mbox{x} $, the constraints being given by the boundary conditions satisfied by $ \textbf{v} $ on the boundary of $ \Omega $. Unfortunately, this result (henceforth referred to as 'Korteweg-Helmholtz' principle') does not apply straightforwardly to our $ \Lambda_{i} \gg 1$, $ Ha \gg 1 $ plasmoid. 

\paragraph{Joule heating.} The amount $ P_{J} $ of heat produced per unit volume and time via Joule heating is

\begin{equation} \label{PJ}
P_{J} = \left( \textbf{E} + \textbf{v} \wedge \textbf{B} \right) \cdot \textbf{j}
\end{equation}

where $ \textbf{E} $ is the electric field - see e.g. eq. XIII.35 of \cite{DeGrootMazur}. Electrons respond more quickly to $ \textbf{E} $ than ions as $ m_{e} \ll m_{i} $, and rule Joule heating. Since $ \Lambda_{i} \gg 1 $, Ohm's law reads \cite{NRL} \cite{Boozer}:

\begin{equation} \label{OHM}
\textbf{E} + \textbf{v} \wedge \textbf{B} - 
\eta_{//} \textbf{j}_{//} - 
\eta_{\perp} \textbf{j}_{\perp} -
\dfrac{\textbf{j} \wedge \textbf{B}}{e n_{e}} +
\dfrac{\nabla p_{e}}{e n_{e}} -
\dfrac{\textbf{R}_{T}}{e n_{e}} -
\textbf{R}_{f} = 0 
\end{equation}

where $ e = 1.6 \cdot 10^{-19} $ C, $ \textbf{j}_{//} \equiv \left( \textbf{b} \cdot \textbf{j}\right) \textbf{b} $, 
$ \textbf{j}_{\perp} \equiv \textbf{b} \wedge \left(\textbf{j} \wedge \textbf{b}\right) $,
 and 
$ \eta_{//} $, $ \eta_{\perp} $, $ p_{e} $, $ \textbf{R}_{T} $ and $ \textbf{R}_{f} $ are the parallel resistivity, the perpendicular resistivity, the partial pressure of electrons, a term $ \propto \nabla T $ and a term responsible for the fluctuations of $ \textbf{B} $ respectively. Both $ \eta_{//} $ and $ \eta_{\perp} $ depend on $ T $, e.g. in the classical treatment of Spitzer. Together with Ampère's law 

\begin{equation} \label{AMPERE}
\textbf{j} = \mu_{0}^{-1} \nabla \wedge \textbf{B} 
\end{equation}

(valid in the non-relativistic limit, with $ \mu_{0} = 4 \cdot \pi \cdot 10^{-7} T \cdot A \cdot m$), Faraday's law

\begin{equation} \label{FARADAY}
\nabla \wedge \textbf{E} + \dfrac{\partial \textbf{B}}{\partial t} = 0
\end{equation}

and Gauss' law of magnetism

\begin{equation} \label{GAUSS}
\nabla \cdot \textbf{B} = 0
\end{equation}

equations \eqref{NS}, \eqref{incompressibility} and \eqref{OHM} allow complete description of our fully ionized, pure-hydrogen plasmoid with the help of the quantities $ \textbf{v} $ and $ \textbf{B} $ only, once the equations of state and the values of $ \rho $, $ \eta_{//} $, $ \eta_{\perp} $, the $ \eta_{\left( V \right) ijkl} $'s, the temperatures and the distance of the plasmoid from the Sun (hence $ \nabla \phi_{g}$) are known. In particular, \eqref{PJ} and \eqref{OHM} allow explicit computation of $ P_{J} $. The contributions of $ \eta_{//} $ and $ \eta_{\perp} $ to $ P_{J} $ add up to $ \textbf{j} \cdot \bm{\eta}_{\left( J \right)} \cdot \textbf{j} $ where $ \bm{\eta}_{\left( J \right)} \equiv \eta_{//} \textbf{b} \textbf{b} + \eta_{\perp} \left( \textbf{1} - \textbf{b} \textbf{b} \right)$; 
note that $ \eta_{\left( J \right) ij} = \eta_{\left( J \right) ji} $. 
The contribution of $ \frac{\textbf{j} \wedge \textbf{B}}{e n_{e}} $ to $ P_{J} $ is 
$ \textbf{j} \cdot \frac{\textbf{j} \wedge \textbf{B}}{e n_{e}} = 0$. 
Both $ \frac{\nabla p_{e}}{e n_{e}} $ and 
$ \frac{\textbf{R}_{T}}{e n_{e}} $ provide also a vanishing contribution, in the limit of vanishing gradients of $ T $ and $ p_{e} \propto n_{e} T $. As for $ \textbf{R}_{f} $, the amplitude of magnetic fluctuations it takes into account is $ \propto \left( \vert \textbf{k} \vert L \right)^{-1} $ where $ \textbf{k} $ is the wavenumber of the fluctuations \cite{Haines}; if $ \textbf{R}_{f} = 0 $ then \eqref{OHM} reduces to the well-known case discussed in \cite{NRL}. In the following we assume $ \vert \textbf{k} \vert L \gg 1 $, then we neglect $ \textbf{R}_{f} $ (for instance, large-$ \vert \textbf{k} \vert $ modes play a crucial role at the birth of a plasmoid in the ionosphere \cite{Birn}). This implies that Joule heating is adequately described with the help of $ \bm{\eta} $, in agreement with \cite{Boozer}: equation \eqref{PJ} reduces to:

\begin{equation} \label{PJoule}
P_{J} = \textbf{j} \cdot \bm{\eta}_{\left( J \right)} \cdot \textbf{j}
\end{equation}

For future reference, here we recall that - as discussed e.g. in \cite{Herrmann}, \cite{Jaynes} and Probl. 3 Sec. 21 of Ref. \cite{Landau} - Kirchhoff has shown that a steady state of an electric conductor at rest (magnetic Reynolds' number $ Re_{m} = 0$) which fills a fixed volume $ \Omega $ and where \eqref{PJoule} holds with $ \eta_{//} = \eta_{\perp} $ and $ \nabla \eta_{//} = 0$ corresponds to a constrained minimum of the total Joule power $ \int_{\Omega} P_{J} d^3\mbox{x} $, the constraint being given by the conservation of electric charge (which in steady state reads $ \nabla \cdot \textbf{j} = 0$, as it turns out by taking the divergence of both sides of \eqref{AMPERE}). Unfortunately, this result (henceforth referred to as 'Kirchhoff's principle') does not apply straightforwardly to our plasmoid, where $ Re_{m} \approx O \left( S \right) \gg 1 $ at all times and $ \eta_{//} \neq \eta_{\perp} $ (even if $ \nabla \eta_{//} = 0 $ is satisfied as $ \eta_{//} = \eta_{//} \left( T \right) $ and $ \nabla T = 0$). 

\paragraph{An useful lemma.} Let us introduce the total density $ P_{h} \equiv P_{v} + P_{J}$ of dissipated power. As a matter of principle, equations \eqref{PV} and \eqref{PJoule} allow us to write $ P_{h} $ in terms of $ \textbf{v} $ and $ \textbf{B} $ only. We show in Appendix \ref{QUAL} that if we apply to a $ \beta \ll 1, \Lambda_{i} \gg 1$ plasmoid satisfying \eqref{NS}-\eqref{PJoule} and contained within a fixed region $ \Omega $ of space a slow, $ \vert \textbf{k} \vert L \gg 1 $ perturbation of $ \textbf{v} $ and $ \textbf{B} $ with identically vanishing $ \frac{\partial \textbf{v}}{\partial t} $ and $ \frac{\partial \textbf{B}}{\partial t} $ on the boundary of $ \Omega $, then the following relationship holds:

\begin{equation} \label{LJAPUNOV}
\dfrac{\partial}{\partial t} \left( \int_{\Omega} P_{h} d^3\mbox{x} \right) \leq 0
\quad \left( = 0 \quad \mbox{in steady state only} \right) \quad ; \quad
T = T_{b} 
\end{equation}

The word 'slow' is given an exact meaning in the Appendix. Remarkably, the symmetries in  $ \bm{\sigma} $ and $ \bm{\eta} $ ensure that both $ P_{v} $, $ P_{J} $, $ P_{h} $ and the total dissipated power $ \int_{\Omega} P_{h} d^3\mbox{x} $ are $ \geq 0 $. Inequality \eqref{LJAPUNOV} implies that a necessary condition for the stability of a steady state is that: 

\begin{equation} \label{variational}
\int_{\Omega} P_{h} d^3\mbox{x} = \min \quad ; \quad T = T_{b}
\end{equation}

further constraints being provided by the relevant equations of motion listed above, i.e. \eqref{NS}, \eqref{incompressibility}, \eqref{AMPERE} and \eqref{GAUSS}. In other words, the total dissipated power behaves as a Ljapunov function. Physically, should a steady state violate \eqref{variational}, the inequality \eqref{LJAPUNOV} would allow a slow perturbation to lead the system farther and farther away from the initial state; a lower bound exixts as $ \int_{\Omega} d^3\mbox{x} P_{h} \geq 0 $. Steady states solve the equations of motion with vanishing $ \frac{\partial}{\partial t} $; stable steady states solve also \eqref{variational}. 

It seems that \eqref{variational} embeds both Korteweg-Helmholtz' and Kirchhoff's principles as particular cases; indeed, it is only the simultaneous validity of $ \beta \ll 1 , \nabla T = 0, \Lambda_{i} \gg 1$ and $ \vert \textbf{k} \vert L \gg 1 $  that makes it possible to overcome the fact that neither advection nor Lorenz force are negligible, in contrast with the assumptions of the original proofs of the principles quoted above. Physically, the smallness of both $ \beta $ and $ \vert \nabla T \vert $ at all times during relaxation implies that both the resistive decay of magnetic field and the viscous decay of the flow pattern leads to no relevant growth of internal energy during the relaxation. In turn, such negligible growth triggers no relevant diamagnetic motion and no further development of the perturbation, so that the amplitude of the fields actually decreases because of dissipation and the same occurs to $ \int_{\Omega} P_{h} d^3\mbox{x} $: basically, the heat produced by decay is assumed to be entirely lost. Neither collisions ($ \Lambda_{i} \gg 1 $) nor magnetic fluctuations ($ \vert \textbf{k} \vert L \gg 1 $) are strong enough, however, to destroy the ordered structure of the relaxed state we are going to describe in the following Section. This limitations make sense as \eqref{variational} is just a necessary condition for the stability of the plasmoid. 

\paragraph{Non-vanishing $ \nabla T $.} By far, the ansatz $ \nabla T = 0 $ is the most unphysical assumption underlying our result \eqref{variational}. (The assumption of equal ion and electron temperature is also unrealistic, but can be easily dropped with straightforward algebra in the following). In order to get rid of it, we take advantage of a lemma of variational calculus, the reciprocity principle for isoperimetric problems -see Sec. IX.3 of \cite{Elsgolts}. This lemma ensures that the solution of the variational problem \eqref{variational} is also the solution of the ('reciprocal') variational problem $ T = \max $ with the constraint of given amount $ W $ of total dissipated power $ \int_{\Omega} P_{h} d^3\mbox{x} $ (all other constraints being unchanged, here and below). This is equivalent to $ T^{-1} = \min ; \int_{\Omega} P_{h} d^3\mbox{x} = W$, which in turn is equivalent to $ \frac{1}{T}\int_{\Omega} P_{h} d^3\mbox{x} = \min ; \int_{\Omega} P_{h} d^3\mbox{x} = W $. Since $ \nabla T = 0 $, we can rewrite the reciprocal version of \eqref{variational} as:

\begin{equation} \label{reciprocal}
\int_{\Omega} \dfrac{P_{h}}{T} d^3\mbox{x} = \min \quad ; \quad \int_{\Omega} P_{h} d^3\mbox{x} = W
\end{equation}

According to \eqref{variational}-\eqref{reciprocal}, and provided that all other constraints provided by the equations of motion are satisfied, if $ \nabla T = 0 $ then stability depends on $ P_{h} $ and $ T $ only. 

Now, when it comes to $ \nabla T \neq 0 $ we may expect that the relationship between $ P_{h} $ and $ T $ in stable plasmoids is somehow modified. However, we have seen that it is reasonable to assume a self-similar evolution for $ n_{e} $, $ T $ and $ \textbf{B} $ in the initial period of plasmoid life at least. In this case, \eqref{PV} and \eqref{PJoule} ensure that the dependence of both $ P_{h} $ and $ T $ on $ \textbf{x} $ is the same at all times; once such dependence is known, both $ W $ and the effective temperature $ T_{eff} \equiv \frac{\int_{\Omega} P_{h} d^3\mbox{x}}{\int_{\Omega} \frac{P_{h}}{T} d^3\mbox{x}} $ encompass all information concerning $ P_{h} $ and $ T $. 

As far as the quantities appearing in \eqref{variational}-\eqref{reciprocal} are involved, a stable (if any), $ \nabla T \neq 0 $ plasmoid with given values of $ T_{eff} $ and $ W $ is indistinguishable from a stable, $ \nabla T = 0 $ plasmoid with given values of $ T_{b} = T_{eff} $ and the same value of $ W $  . 
In other words, as far as we are dealing with a self-similar succession of relaxed, quasi-steady-states with given profiles of $ n_{e} $, $ T $ and $ \textbf{B} $, the problem of stability of a $ \nabla T \neq 0 $ plasmoid is equivalent to the corresponding problem of a $ \nabla T = 0 $ plasmoid with $ T \left( \textbf{x} \right) = T_{eff} $ everywhere, and with the same $ W $. Roughly speaking, self-similarity 'freezes' the additional degree of freedom made available by $ \nabla T \neq 0 $, so that the problem of stability reduces basically to the $ \nabla T = 0 $ case discussed above. (In Ref. \cite{PRE2010} a derivation of \eqref{reciprocal} is presented starting from the assumption of local thermodynamic equilibrium, which does not apply here). With this proviso, we are going to take advantage of \eqref{variational} in the following.

\section{Double Beltrami}
\label{SEC4}

Steady-state solutions of the Euler-Lagrange equations of the variational principle \eqref{variational} constrained by \eqref{NS}, \eqref{incompressibility}, \eqref{AMPERE} and \eqref{GAUSS} include the fields $ \textbf{v} $ and $ \textbf{B} $ which solve the following relationships \cite{Di Vita00}:

\begin{equation} \label{DBv}
\nabla \wedge \textbf{v} = r \textbf{v} + w \textbf{B} + \nabla \varphi
\end{equation}

\begin{equation} \label{DBB}
\nabla \wedge \textbf{B} = l \textbf{v} + g \textbf{B} + \nabla \chi
\end{equation}

where $ r, w, l $ and $ g $ are constant quantities (their actual values are not relevant here); moreover, $ \varphi $ and $ \chi $ are harmonic fields which represent the interaction of the system with the external world. 

In the relevant case where this interaction is relatively weak (e.g. in comparison with the magnetic interaction among currents internal to the system) a number of relevant results are proven in \cite{Di Vita00} \cite{Di Vita01} \cite{Kagan} \cite{Ohsaki}. To start with, it is useful to write 
$ \vert \nabla \wedge \textbf{v} \vert^{-1} \vert \nabla \varphi \vert \approx \vert \nabla \wedge \textbf{B} \vert^{-1} \vert \nabla \chi \vert \approx O \left( \varepsilon \right), 0 < \varepsilon \ll 1 $. 

Firstly, it has been shown \cite{Di Vita00} that 
$ r + g \approx rg - wl \approx \vert \textbf{j} \wedge \textbf{B} \vert \approx O \left( \varepsilon \right) $. This result is invoked below. Physically, it means that the results of this Section are reasonable at least as far as $  O \left( \varepsilon \right) \approx \vert \textbf{j} \wedge \textbf{B} \vert \approx O \left( \beta \right) $, i.e. in our initial, low-$ \beta $ period of plasmoid lifetime. 
Thus, the above scalings $ \nabla \varphi \approx O \left( \varepsilon \right) $ and $ \nabla \chi \approx O \left( \varepsilon \right) $ agree with the virial theorem of MHD, which states that no magnetically confined plasma exists without interaction with external currents. Accordingly, we neglect terms $ \approx O \left( \varepsilon \right) $ altogether in the following.

Secondly, the solutions of \eqref{DBv}-\eqref{DBB} solve also the Euler-Lagrange equations of Turner’s variational principle \cite{Turner} which minimizes the sum $ E \equiv E_{K} + E_{M} $ of kinetic energy 
$ E_{K} \equiv \int_{\Omega} \frac{\rho \vert \textbf{v} \vert^2 }{2} d^3\mbox{x}$ and magnetic energy 
$ E_{M} \equiv \int_{\Omega} \frac{\vert \textbf{B} \vert^2 }{2 \mu_{0}} d^3\mbox{x}$ with the constraints of fixed magnetic helicity 
$ K \equiv \int_{\Omega} \textbf{A} \cdot \textbf{B} d^3\mbox{x} $ and
generalized helicity 
$ H \equiv \int_{\Omega} \bm{\Omega} \cdot \textbf{V} d^3\mbox{x} $. Here 
$ \bm{\Omega} = \frac{e}{m_{i}}\textbf{B} + \nabla \wedge \textbf{v} $ and $ e = 1.6 \cdot 10^{-19} C$; moreover, 
$ \textbf{A} $ and $ \textbf{V} $ are the vector fields such that 
$ \textbf{B} = \nabla \wedge \textbf{A} $ and 
$ \bm{\Omega} = \nabla \wedge \textbf{V} $ respectively. 
Turner’s variational principle has been applied to CMEs and to plasmoids in the lab in Refs. \cite{Kagan}, \cite{Ohsaki} and in Refs. \cite{Auluck1}, \cite{Auluck2} and \cite{Di Vita02} respectively. Historically, it has been postulated \cite{Turner} as a generalization to Hall MHD of Taylor’s variational principle of minimization of $ E_{M} $ with the constraint of fixed $ K $ \cite{Taylor}. Admittedly, this postulate has been put in doubt in Ref. \cite{Yoshida}; here, however, it follows from \eqref{variational} and is not a postulate anymore. For future reference, we recall that the $ \textbf{B} $ which solves Taylor's principle satisfies Beltrami equation $ \nabla \wedge \textbf{B} = \lambda \textbf{B} $ with $ \nabla \lambda = 0 $ and linear size $ \lambda^{-1} = \left( 2 \mu_{0} E_{M} \right)^{-1} K $, hence $ \textbf{j} \wedge \textbf{B} = 0 $ exactly. This equation is often invoked when describing plasmas both in lab (reversed field pinches \cite{Bodin}, spheromaks \cite{Bellan}) and in space (CMEs \cite{WANG} \cite{Low}, jets like in NGC6251 \cite{Choudhuri}). Remarkably, both Turner's and Taylor's principle neglect internal energy; this is reasonable in the initial, low-$ \beta $ period of plasmoid lifetime at least.

Thirdly, equations \eqref{DBv}-\eqref{DBB} lead to the following relationship for $ \textbf{B} $:

\begin{equation} \label{DB}
\left( \nabla \wedge - \lambda_{1} \right)
\left( \nabla \wedge - \lambda_{2} \right) \textbf{B} = 0
\end{equation}

where 
$ \lambda_{1} \approx \left( r + g \right)^{-1} \left( r g - w l \right) \approx O \left( \varepsilon^0 \right) \gg \lambda_{2} \approx r + g \approx O \left( \varepsilon^1 \right)$ and we have taken into account that 
$ r + g \approx rg - wl \approx O \left( \varepsilon \right) $, i.e. $ \textbf{B} $ exhibits both a small-scale structure with typical length $ \lambda_{1}^{-1} $ and a large-scale structure with typical length $ \lambda_{2}^{-1} \gg \lambda_{1}^{-1} $. This separation between large and small spatial scales had been postulated with no further proof in \cite{Ohsaki} and turns out to be justified by observations of CMEs in \cite{Kagan}; here, it follows from \eqref{variational}. As a result, $ \textbf{B} $ is a superposition \cite{Yoshida} of two Taylor fields with different typical lengths, hence the name ‘Double Beltrami’ (DB) for the field $ \textbf{B} $ which solves \eqref{DB} \cite{Kagan}. (A similar expression holds for $ \textbf{v} $). 

Fourthly, if $ \lambda_{2} $ exceeds a threshold value $ \lambda_{c} $ then the system undergoes a severe reorganization: the contribution of the small-scale Taylor field to the DB state vanishes, and the system reduces abruptly to a single Beltrami (i.e., Taylor) state $ \nabla \wedge \textbf{B} = \lambda_{c} \textbf{B} $, with $ \textbf{v} $ parallel to $ \textbf{B} $. (The word 'abruptly' is given a precise meaning below). In the treatment of Ref. \cite{Ohsaki} the cases $ \lambda_{2} < \lambda_{c} $, $ \lambda_{2} = \lambda_{c} $ and $ \lambda_{2} > \lambda_{c} $ correspond to a solar arcade magnetic field structure resembling interacting coronal loops, to a solar eruption and to an ejected plasmoid respectively; the velocity of the latter is predicted in equation (21) of \cite{Kagan}. In this model, it is at $ \lambda_{2} = \lambda_{c} $ that lines of force of the solar arcade quickly reconnect into a low arcade of loops, leaving a helix of magnetic field unconnected to the rest of the arcade. At the transition $ E $ and $ E_{M} $ take the values $ E_{c} $ and $ E_{M c} $ respectively, with $ E_{c} \approx 2 E_{M c} $ (see final lines of §2.3 of \cite{Kagan}), and equation (16) of \cite{Ohsaki} links $ \lambda_{c}, E_{M c} $ and the value $ K_{c} $ of $ K $ at the transition, i.e. 
$ \frac{c}{\omega_{pi}} \lambda_{c} \approx \left( 2 \mu_{0} E_{c} \right) \left( 2 K_{c} \right)^{-1} \approx \left( 2 \mu_{0} E_{M c} \right) K_{c}^{-1} $ 
in dimensional units, where $ \omega_{pi} $ and $ c $ are the ion plasma frequency and the speed of light in vacuum respectively. For given $ n_{e} = n_{i} \propto \omega_{pi}^{2} $, transition to Taylor’s state occurs therefore at a threshold value of $ K \left( 2 \mu_{0} E_{M} \right)^{-1} $. 

Each of these results finds its own correspondent in the lab. To start with, spontaneous formation of vortex-like structures in Hall MHD occurs \cite{Stenzel} on a time-scale $ \ll $ the period of a ion Larmor orbit: thus, the word ‘abruptly’ above is given a precise meaning. Remarkably, such structures have a typical linear size $ \approx \frac{c}{\omega_{pi}} $ \cite{Witalis2}. Furthermore, it has been shown \cite{Di Vita01} that \eqref{variational} links $ K \left( 2 \mu_{0} E_{M} \right)^{-1} $ and $ Ha $. Transition to Taylor’s state occurs whenever:

\begin{equation} \label{thr1}
K \left( 2 \mu_{0} E_{M} \right)^{-1} > K_{c} \left( 2 \mu_{0} E_{M c} \right)^{-1} = 
0.3 \cdot \dfrac{c}{\omega_{pe}} \cdot \max{\left(1, \Lambda_{i}\right)}
\end{equation}

where $ \omega_{pe} $ is the electron plasma frequency. This inequality means that if the magnetic helicity content is too large then the system changes to a Taylor-state. (Remember that we are speaking of plasmas where the impact of external, possibly stabilizing fields is neglected). Remarkably, resistive decay of $ K $ is usually much slower than the resistive decay of $ E_{M} $ - see e.g. the numerical results of  \cite{Dasgupta} concerning relaxation of a weakly resistive, inviscid plasma towards a state described by \eqref{variational}; these results seem to confirm previous suggestions of Ref. \cite{Dasgupta2}. Moreover, $ \Lambda_{i} \propto \vert \textbf{B} \vert \propto \sqrt{ E_{M}}$. Accordingly, the more twisted the magnetic field lines (i.e., the larger $ K $), the more likely the occurrence of a solar eruption during the evolution of a flux rope. This picture of a solar eruption as a result of the instability of a twisted flux rope is not in contrast with recent observations \cite{Amari}. Moreover, \eqref{thr1} is satisfied if and only if:

\begin{equation} \label{thr2}
Ha > Ha_{c} = 50 \cdot \sqrt{I(MA)} \cdot T_{e}(KeV)^{-1/4} \max{\left(1, \sqrt{\Lambda_{i}}\right)}
\end{equation}

where $ I \approx \frac{\vert \textbf{B} \vert \cdot L}{\mu_{0}} $ is the typical amount of electric current flowing across the plasma. Relationships \eqref{thr1}-\eqref{thr2} describe both the transition from single helicity states to multiple helicity states \cite{Escande} in a Reversed Field Pinch \cite{Bodin2} and the transition \cite{Di Vita00} from filaments \cite{Herold} to hot spots \cite{Bernard} in the pinch of a Dense Plasma Focus (DPF) \cite{Soto1}; both transitions correspond to a change in geometry, in qualitative analogy to what is predicted at the birth of a plasmoid in the ionosphere \cite{Birn}. According to \cite{Soto2}, filaments in the DPF pinch behave as relaxed stated described by Turner's principle; moreover, their typical size is $ \propto \frac{c}{\omega_{pi}} $, in qualitative agreement with the results of \cite{Di Vita00}, \cite{Stenzel} and \cite{Witalis2} quoted above. If $ Ha < Ha_{c} $ the DB state has a filamentary structure; if $ Ha > Ha_{c} $ it is similar to a Taylor state. It is easy to see that \eqref{thr2} is satisfied for the parameters listed in Sec. \ref{SEC2}, i.e. our plasmoid is actually a Taylor state, as $ Ha_{c \left(1 A.U.\right)} \approx 10^6$ and $ Ha_{c \left(init\right)} \approx 5 \cdot 10^6$. (Admittedly, this conclusion is only marginally valid far from the Sun; the issue is further discussed in Sec. \ref{SECMERGING} ). This fact provides a further justification of the force-free expression for $ \textbf{B} $ e.g. in \cite{WANG}; now, however, it is not an ansatz anymore, as it follows from \eqref{variational}. 

Some final remarks concerning turbulence. We have adopted everywhere the familiar approach to viscosity ruled by collisions \cite{NRL}. This makes sense as far as the ion-ion collision time $ \tau_{ii} $ is short enough. Indeed, its values are $ \approx 5 s $ and $ \approx 10^{7} s $ near the Sun and at 1 A.U. respectively, so our assumption makes sense in the initial phase of the plasmoid life only. Generally speaking, relaxation to Taylor's state requires some level of turbulence \cite{Taylor}; turbulence-induced ion heating is therefore possible beyond collisional viscous effects, and $ P_{v} $ may be affected. According to \cite{Haines2}, this ion heating is fed by $ \vert \textbf{k} \vert L \approx 10^2 $ MHD instabilities with wavenumber $ \textbf{k} $ whenever 
$ \frac{ \rho \cdot c_{sA} }
{
\vert \textbf{k} \vert 
\cdot 
\left[ 
\frac{1}{3}
\eta_{\left(V 0 \right)}
+ 
\eta_{\left(V 0 \bot \right)}
\right] 
}
 \approx 1 $
 where $ c_{s} $ and 
 $ \eta_{\left(V 0 \bot \right)} \approx 
 \eta_{\left(V 0 \right)} \cdot \Lambda_{i}^{-2} $ are the speed of sound and the perpendicular viscosity coefficient respectively. 
Far from the Sun, this reduces to $ Re \approx 100 $, which is violated as $ Re_{\left(1 A.U.\right)} \approx 1 $, i.e. 
there are just too many collisions for efficient turbulent ion heating: ions are heated mainly by collisions, and our treatment is self-consistent. 

\section{The scattering}
\label{SEC5}

We have dealt with isolated plasmoids so far. When dealing with mutually interacting plasmoids, differences between plasmoids in laboratory and in space become relevant. In the lab, prolate, $ Re_{m} \ll 1 $ plasmoids are often observed \cite{Bostick} \cite{Jakubowski} whose evolution conserves neither $ K $ nor $ H $ - see Ref. \cite{Di Vita02} and Refs. therein. In contrast, during the flight across space of our plasmoid from the Sun towards the region where scattering occurs, we may safely write: 

\begin{equation} \label{Kcons}
K = \mbox{const.} 
\end{equation}

\begin{equation} \label{Hcons}
H = \mbox{const.} 
\end{equation}

These results are far from trivial, as the evolution of the plasmoid is not necessarily self-similar as the distance from the Sun increases; they are justified as both $ Re $ and $ Re_{m} $ are $ > 1 $ during the flight at all times. Since the (resistive) decay of $ K $ and the (viscous + resistive) decay of $ H $ are much slower than the decay of $ E_{K} $ and $ E_{M} $ respectively, dissipation may as well raise the insofar neglected internal energy at the expense of $ E $ during the flight (thus raising $ \beta $).

Remarkably, moreover, it seems reasonable to assume that $ K $ and $ H $ are conserved even during the scattering, with the proviso that (with a slight misuse of notation, admittedly) by $ K $ and $ H $ we refer to the total amount of magnetic helicity and generalized helicity respectively. In fact, an upper bound on the collisional diffusion coefficient in our magnetized plasmoid is $ \frac{k_{B} T}{m_{i}} \cdot \tau_{ii} $ (the actual value may be lower, as $ \Lambda_{i} > 1$; $ k_{B} $ is Boltzmann's constant). Accordingly, an upper bound on the typical diffusion length at the scattering is provided precisely by $ Re_{\left(1 A.U.\right)} \approx 1 $, i.e. $ \sqrt{\Delta t \cdot \frac{k_{B} T}{m_{i}} \cdot \tau_{ii}} \approx L_{\left(1 A.U.\right)} $. Similar arguments hold for magnetic diffusion. This means that the scattering is so short that resistive and viscous decay processes leave global invariant quantities like $ K $ \cite{Taylor} and $ H $ \cite{Turner} unaffected.

Furthermore, total mass $ M \equiv \int_{\Omega} d^3\mbox{x} \rho $ is conserved during the scattering:

\begin{equation} \label{Mcons}
M = \mbox{const.} 
\end{equation}

As for energy, of course the total energy of the system - the sum of gravitational energy, internal energy $ \approx \beta \cdot E_{M} $ and of $ E $ - is conserved. In particular, plasmoid-plasmoid gravitational interaction is negligible, when compared to electromagnetic interaction. Moreover, we are allowed to neglect variations of the plasmoid potential energy in the solar gravitational field during scattering, because $ \left[ v_{0} \cdot \Delta t \cdot \phi_{g}^{-1} \cdot \vert \nabla \phi_{g} \vert \right]_{1 A.U.} \ll 1 $. Generally speaking, we cannot neglect internal energy, as $ \beta_{\left(1 A.U.\right)} \approx 0.4 $. For the moment, we assume that not only $ K $ and $ H $, but also $ E $ is conserved:

\begin{equation} \label{Econs}
E = \mbox{const.} 
\end{equation}

Conservation of total energy makes \eqref{Econs} to be equivalent to separate conservation of internal energy and of $ E $, i.e. they do not transform into each other during the scattering. 
This is e.g. possible in weakly dissipating ($ S_{\left(1 A.U.\right)} \approx 2 \cdot 10^{13}, Ha_{\left(1 A.U.\right)} \approx 10^{6} $), incompressible systems, hence free from shock waves. We discuss the impact of dissipation on scattering in Sec. \ref{SECMERGING}, and hint briefly at the role of shock waves at the end of Appendix \ref{QUAL3}. 

Admittedly, conversion of internal energy into kinetic energy has been reported \cite{TURNON} \cite{COULD}; indeed, even if the velocities in the center-of-mass system are $ < c_{sA} $, compressibility is not so quite obviously negligible, as $ \beta_{\left(1 A.U.\right)} \approx 0.4 $. However, this is likely to leave our discussion of super-elastic collisions unaffected. Indeed, Figs. 1a, 1d and 1e of \cite{TURNON} and Fig. 3 of \cite{COULD} suggest that both $ E_{M} $ and internal energy $ \propto \beta \cdot E_{M} $ decrease as $ E_{K} $ increase in super-elastic collisions; we may suggest that the impact of dissipation on $ \beta $ is too weak, so that we are allowed to limit ourselves to take just into account $ E_{M} $, as internal energy will follow in all cases. In contrast, Figs. 1b and 1c of \cite{TURNON} suggest that inelastic scattering reduces $ E_{K} $ and raises $ \beta $, with substantial growth of internal energy. This suggests discussion of two extreme, opposite classes of scattering problems. In the first class, \eqref{Econs} is violated and we allow dissipation-induced merging of plasmoids, i.e. the kinetic energy goes to zero in the center-of-mass system. 
In the second class, both \eqref{Kcons}, \eqref{Hcons}, \eqref{Mcons} and \eqref{Econs} apply, and internal energy plays no independent role. The first class is discussed in Sec. \ref{SECMERGING}. We discuss the second class in the following.

We limit ourselves to a head-on collision in the frame of reference of the center of mass of a system of two plasmoids. Before (after) the collision, the plasmoids (say, 1 and 2) move towards (away from) each other on the same straight line, which is parallel to the $ z $ axis. For simplicity, we assume that 1 and 2 have the same mass, and the typical linear size $ \lambda^{-1} $ of 1 is equal to the corresponding quantity of 2 at all times (we drop the last assumption below). Before (after) the collision, the centers of mass of 1 and 2 move at velocity $ + v_{C} \textbf{z} $ ($ - v_{C} \textbf{z} $) and $ - v_{C} \textbf{z} $ ($ + v_{C} \textbf{z} $) respectively. In this head-on collision, relevant conservation laws are concerned with total mass, energy, magnetic flux and component of momentum along $ \textbf{z} $; these 4 independent relationships correspond to the 4 independent equations \eqref{Kcons}, \eqref{Hcons}, \eqref{Mcons} and \eqref{Econs}.

We denote with $ \textbf{v}_{1}' $ the macroscopic velocity of a small mass element of plasma of 1 in the frame of reference of the center of mass of plasmoid 1. Then, the macroscopic velocity $ \textbf{v}_{1} $ of the small mass element in the frame of reference of the center of mass of the system is $ \textbf{v}_{1} = + v_{C} \textbf{z} + \textbf{v}_{1}' $ ($ \textbf{v}_{1} = - v_{C} \textbf{z} + \textbf{v}_{1}' $) before (after) the collision. Similar notations hold for 2. Of course, $ \textbf{v}_{1} $ ($ \textbf{v}_{2} $) vanishes outside 1 (2). Finally, we write $ \textbf{B} = \textbf{B}_{1} + \textbf{B}_{2} $, where $ \textbf{B}_{1} $ ($ \textbf{B}_{2} $) is the magnetic field produced by electric current flowing across plasmoid 1 (2).  Since both 1 and 2 are in a Taylor state and $ \lambda $ is the same, $ \textbf{B} $ too is in Taylor's state, i.e. spheromak-like with typical linear size $ \lambda^{-1} $. As for 1 (2), therefore, the volume averages 
$ \langle \textbf{B}_{1 \left( 2 \right)} \cdot \textbf{z} \rangle $ vanish 
(here and in the following we denote by $ \langle \emph{a} \rangle $ the volume average of the generic quantity $ \emph{a} $). Since 
$ \textbf{v}_{1}' \wedge \textbf{B}_{1} = 0 $ and 
$ \textbf{v}_{2}' \wedge \textbf{B}_{2} = 0 $ in Taylor-like solutions of Turner's principle \cite{Di Vita02}, we write also 
$ \langle \textbf{v}_{1 \left( 2 \right)}' \cdot \textbf{z} \rangle = 0$. Furthermore, we take $ \nabla \rho = 0 $ for mathematical simplicity. This choice seems reasonable if we allow strong turbulence to enhance transport throughout plasma volume during the scattering. Implicitly, however, here we assume that scattering is so short that turbulence-induced growth of resistivity and viscosity has negligible effects on electrons and ions respectively. Finally, \eqref{Kcons}, \eqref{Hcons}, \eqref{Mcons} and \eqref{Econs} lead to:

\begin{equation} \label{intid}
H - \left( \dfrac{e}{m_{i}} \right)^2 K = 
\int_{\Omega} d^3\mbox{x} \left( \textbf{v}_{1}' \cdot \nabla \wedge \textbf{v}_{1}'  
+ \dfrac{2 e}{m_{i}} \textbf{v}_{1}' \cdot \textbf{B} \right) + \left( 1 \longleftrightarrow 2 \right)
= \mbox{const.}
\end{equation}

\begin{equation} \label{intid2}
E = \dfrac{\lambda K}{2 \mu_{0}} + 
\left( \dfrac{M}{2} \right) 
\left[ v_{C}^2 + 
\langle \vert \textbf{v}_{1}' \vert ^ 2 \rangle + 
\langle \vert \textbf{v}_{2}' \vert ^ 2 \rangle 
\right]
= \mbox{const.}
\end{equation}

Let us investigate the behaviour of each term in \eqref{intid} and \eqref{intid2} under the scaling transformation:

\begin{equation} \label{scaling}
\textbf{x} \rightarrow k \textbf{x}
\end{equation}

We have to ensure that \eqref{scaling} leaves the values of the above discussed quantities unaffected (after all, we cannot create mass, or kinetic energy, or twistedness of the magnetic field lines just by rescaling). Then, we may invoke neither the results of Sec. 3.2 of Ref. \cite{Alfvèn} (which leaves electrostatic potential unaffected, and is therefore not relevant here where no electrostatic potential is considered) nor the Collisional Vlasov High Beta Scaling \cite{Di Vita00} \cite{Di Vita02} \cite{Connor} (which applies to plasmas with $ Re_{m} < 1 $). The Jacobian of \eqref{scaling} is $ k^3 $,  i.e. both the volume $ \approx L^3 $ of a plasmoid and the volume element $ d^3\mbox{x} $ in the volume integrals above undergoes multiplication by $ k^3 $. Then, \eqref{Mcons} implies $ M \rightarrow k^0 M $, so that $ \rho \rightarrow k^{-3} \rho $. Since $ \textbf{B} = \nabla \wedge \textbf{A} $, \eqref{Kcons} implies $ K \rightarrow k^0 K $, $ E_{M} \rightarrow k^{-1} E_{M} $, $ \textbf{B} \rightarrow k^{-2} \textbf{B} $ and $ \textbf{A} \rightarrow k^{-1} \textbf{A} $. Equations \eqref{Kcons}, \eqref{Hcons} and \eqref{intid} imply that the R.H.S. of \eqref{intid} scales as $ k^0 $. But 
$ \int_{\Omega} d^3\mbox{x} \left( \textbf{v}_{1}' \cdot \nabla \wedge \textbf{v}_{1}' \right) $ and $ \frac{2 e}{m_{i}} \left( \textbf{v}_{1}' \cdot \textbf{B} \right) $ scale as $ k^2 \vert \textbf{v}_{1}' \vert ^2 $ and $ k \vert \textbf{v}_{1}' \vert $ respectively. Then, \eqref{intid} is satisfied only if $ k \vert \textbf{v}_{1}' \vert $ is constant, i.e. $ \vert \textbf{v}_{1}' \vert \propto k^{-1} $. The same holds for $ \textbf{v}_{2}' $, because of the symmetry of the R.H.S. of \eqref{intid}. Finally, \eqref{Kcons}, \eqref{Mcons}, \eqref{Econs} and \eqref{intid2} give:

\begin{equation} \label{v0k}
v_{C}^2 = \dfrac{2}{M} \left[ E - \dfrac{E_{M \left( k = 1 \right)}}{k} \right] - 
\dfrac{\left[ \langle \vert \textbf{v}_{1}' \vert ^ 2 \rangle + 
\langle \vert \textbf{v}_{2}' \vert ^ 2 \rangle \right]_{\left( k = 1 \right)}}{k^2}
\end{equation}

where we have taken into account that $ \frac{K \lambda}{2 \mu_{0}} = E_{M} $ in Taylor's states. Equation \eqref{v0k} shows that $ \vert v_{C} \vert $ is an increasing function of $ k $, i.e. any process which makes a plasmoid larger (i.e., which raises its volume) raises also the linear kinetic energy of its center of mass. Crucially, the result does not depend on the detailed physical mechanism of the enlarging process.

We have assumed that $ \lambda $ is the same in 1 and 2. Qualitatively, it is easy to see that our results are not heavily affected if the values of $ \lambda $ in 1 and 2 are different, as far as \eqref{thr1} is satisfied. On the contrary, let us take e.g. the opposite case of vanishing total magnetic helicity $ K $ (we refer to the results reported in \cite{Ono}, where plasmoids 1 and 2 have opposite helicities). In this particular case, if one stable plasmoid turns out to be the final outcome of the collision, then $ \nabla p $ is no more zero, i.e. the resulting plasmoid is a non-Taylor DB structure \cite{MahajanYoshida}.

Now, let us focus our attention on one plasmoid, say 1, with no loss of generality. As 2 approaches, the magnetic flux across 1 changes. This change induces Faraday currents across 1, and these currents lead to Joule heating of 1. As usual by now, we neglect any change in $ p $ across 1 during the scattering. Since the internal energy density is $ \frac{3 p}{2} $, the amount $ \Delta Q = \Delta t \cdot \eta_{//}^{-1} \vert \textbf{E} \vert ^ 2 \cdot L^3 $ of heat produced by Joule heating in a time $ \Delta t $ is equal to $ \frac{3 p}{2} \cdot \Delta V $, where $ \Delta V $ is the increase in the volume $ \approx L^3 $ of 1 and $ \textbf{E} \approx v_{C} \vert \textbf{B} \vert $. Correspondingly, $ k = \sqrt[3]{\frac{L^3 + \Delta V}{L^3}} \approx \sqrt[3]{1 + \frac{Re_{m}}{\beta}} \gg 1 $, as far as $ Re_{m} \approx \mu_{0} \cdot \eta_{//}^{-1} \cdot \Delta t \cdot v_{C}^2 $. Similar arguments apply to 2. 

Admittedly, this is just a qualitative argument. In particular, values $ \gg 1 $ of $ k $ are in contradiction with \eqref{Econs}, which relies basically on the assumption of negligible dissipative decay of $ E $ during the scattering (i.e., of very short $ \Delta t $ and, correspondingly, negligible $ \Delta V $). Indeed, the argument is weakened in two ways at least. Firstly, the assumption of Spitzer resistivity is questionable, as the electric field $ \vert \textbf{E} \vert \approx  \vert v_{C} \vert \cdot \vert \textbf{B} \vert \approx 5 \cdot 10^{-3} \frac{V}{m}$ is $ \gg $ the far-from-the-Sun value $ 2.5 \cdot 10^{-10} \frac{V}{m}$ of the Dreicer field $ \frac{ 5 \cdot 10^{-16} \cdot n_{e}\left(m^{-3} \right)}{T \left(eV \right)} \frac{V}{m}$ required for runaway of electrons at energy $ k_{B} T $ \cite{Miyamoto}. Consequently, wave-particle interactions may occur, which raise the effective resistivity and reduce $ k $. Secondly, the assumption of costant $ p $ is likely to be unphysical; the argument above shows that any finite growth of $ p $ during the scattering decreases $ k $. 

All the same, we may conclude that scattering enlarges plasmoids. According to \eqref{v0k}, therefore, the scattering raises their linear kinetic energy (at the expense of the sum of their magnetic energy), i.e. the scattering is super-elastic. For example, the energy balance discussed in \cite{ShenNature} provides just a 6.6 \% increase in the linear kinetic energy. 

More precise predictions - concerning e.g. the evolution of $ E_{M} $ and $ E_{K} $ during the scattering, the detailed mechanism reponsible for the conversion of magnetic energy into linear kinetic energy, etc. - require thorough analysis of both $ \langle \vert \textbf{v}_{1}' \vert ^ 2 \rangle $ and $ \langle \vert \textbf{v}_{2}' \vert ^ 2 \rangle $, as well as a detailed balance \cite{Cargill} of energy, momentum, etc., a task which lies outside the scope of the present work. 

Finally, the lower the impact velocity, i.e. the initial value $ v_{C \left( k = 1 \right)} $ of $ v_{C} $, the lower the initial value $ E_{K \left( k = 1 \right)} $ of  $ E_{K} $ (namely, the value of the square-bracketed quantity on the R.H.S. of \eqref{v0k} for $ k = 1 $), the larger the relative increase of $ v_{C} $ after the scattering (according to \eqref{v0k}), the larger the relative increase in the linear kinetic energy. We retrieve the anti-correlation suggested in \cite{ShenNature} : in the words of \cite{TURNON}, \textit{the collision with the smaller approaching speed tends to be super-elastic}.

\section{Merging vs. bouncing}
\label{SECMERGING}

So far, we have discussed the ('bouncing') case where the plasmoids move towards (away from) each other before (after) the collision \cite{ShenNature} \cite{Mishra} \cite{Karimabadi}. However, the nature of collision of interacting plasmoids have been investigated by various authors and is found that its regime can range from super-elastic to inelastic \cite{Mishra}. It is even possible that two colliding plasmoids just merge into each other ($ E_{K} \rightarrow 0 $ in the center-of-mass system after collision) \cite{Tajima} \cite{Karlicky} \cite{Karimabadi} \cite{Jelinek} \cite{RHESSI}, rather than bouncing off each other.
Physically, it is reasonable to assume that the choice between bouncing and merging depends on the impact velocity (which we denote by $ v_{C} $ for simplicity below, with no loss of generality). Again, the analogy with elastoplastic collisions - see Fig. 6 in \cite{Magnanimo} - suggests that a dissipation-related criterion exists: if $ v_{C} $ is below (above) a threshold $ v_{thr} $, then collisions lead to bouncing (merging). 
As for plasmoids, Fig. 4.c of \cite{TURNON} seems to confirm such suggestion. Intuitively,  the larger $ v_{C} $, the faster the change of magnetic flux across 1 due to the approach of 2 (and vice-versa), the larger the induced electric field and the corresponding Faraday currents, the stronger the violation of that equation \ref{Econs} during the scattering which underlies our discussion of bouncing in Sec. \ref{SEC5} , the easier the merging. In the opposite limit of low $ v_{C} $, in contrast, bouncing rules - and the lower $ v_{C} $ the more super-elastic the scattering, as shown above. 

However, this simple picture raises a difficult question. A discussion of the dependence of the final outcome of scattering on the relative velocity requires detailed knowledge of momentum balance, which in turn depends on the motion of ions as $ m_{i} \gg m_{e} $. Generally speaking, the global evolution of the system depends on the ion pressure tensor, whose anisotropic and agyrotropic nature is due to the meandering orbits of ions and may not be correctly described in current fluid models \cite{Stanier} \cite{Ng}.
Fully kinetic simulations seem therefore to be useful \cite{Karimabadi} \cite{Karlicky}, Admittedly, however, their utilization in realistic, 3D problems is limited because of the computational resources required. 
We get round this the following way. We try to obtain information from the investigation of Joule heating because Joule heating is ruled by electrons and is therefore affected by the complicated orbits of ions only weakly. The price to be paid is that our discussion is qualitative; then, the operator '$ = $' refers to order-of-magnitude-estimates only in the following of this Section. 

Since we are interested in merging, we are allowed to identify the typical time-scale $ \Delta t = \frac{L}{v_{C}} $ of scattering with the duration of the merging process. Since dissipation rules merging, $ \Delta t $ is also the time over which significant energy release via magnetic reconnection occurs; reconnection allows the large-scale rearrangement of magnetic topology required by merging, and reconnection-induced modification of magnetic field lines affects their magnetic tension, which in turn  drives the motion of the plasmoids \cite{RHESSI}. 

Our discussion is made of two steps. Firstly, we follow step-by-step the simplified treatment of reconnection of Ref. \cite{Cassak}. Finally, we apply it to our merging problem.

Consider two regions of magnetic field coming together and reconnecting. Let $ \textbf{B} $ thread a region of linear size $ R $ along the straight line connecting the centers of the two regions and of linear size $ L_{ext} $ in the direction of the reconnection electric field $ \textbf{E} $. Both $ R $ and $ L_{ext} $ are assumed uniform for simplicity. Then, the amount $ \Delta \Phi $ of magnetic flux processed per unit time by reconnection is $ \frac{\Delta \Phi}{\Delta t} $, where $ \Delta \Phi = \vert \textbf{B} \vert R L_{ext} $. (Implicitly, this model assumes $ \textbf{E} \cdot \textbf{B} = 0 $, which agrees with \eqref{Kcons} as far as no amount of magnetic helicity $ K $ is assumed to flow across the boundary of the system of merging plasmoids at any time). According to \eqref{FARADAY}, the voltage $ V = L_{ext} \vert \textbf{E} \vert $ is equal to $ V = \frac{\Delta \Phi}{\Delta t} $, hence $ \vert \textbf{E} \vert = \frac{\vert \textbf{B} \vert R}{\Delta t} $, i.e. $ M_{A} = \frac{R}{c_{A} \Delta t} $ where we have introduced the dimensionless 'global reconnection rate' $ M_{A} \equiv \frac{\vert \textbf{E} \vert}{\vert \textbf{B} \vert c_{A}}$ \cite{Stanier} \cite{Cassak} \cite{Comisso} \cite{Ng} and the Alfvèn velocity $ c_{A} \equiv \frac{\vert \textbf{B} \vert}{\sqrt{\mu_{0} m_{i} n_{i}}}$. Physically, the numerator $ R $ and the denominator $ c_{A} \Delta t $ in $ M_{A} $ are the radial distance of magnetic flux reconnected in the time $ \Delta t $ and the distance that would have been reconnected in the same time if the inflow speed was $ c_{A} $ respectively. This is consistent with the more commonly quoted form of the reconnection rate \cite{RHESSI}, namely $ M_{A} \equiv \frac{v_{in}}{c_{A}} $ where $ v_{in} \equiv \frac{R}{\Delta t} = \frac{\vert \textbf{E}\vert }{\vert \textbf{B} \vert}$ is the absolute value of the inflow speed. The normalizing value of $ \vert \textbf{B} \vert c_{A} $ in $ M_{A} $ is taken asymptotically far from the ('micro-scale') region ruled by magnetic diffusion \cite{Liu0}, rather than immediately upstream of it. 

Now, it comes to the merging. Plasmoids move towards each other along a common direction $ \textbf{z} $. As they meet, and before the merging is completed, we assume magnetic reconnection to start in a micro-scale region near the point of contact; in both plasmoids, the inflow speed of plasma into such region is directed along $ \textbf{z} $. Plasmoids keep on drawing themselves up to each other, thus feeding magnetic reconnection with magnetic flux. 

The difference between bouncing and merging is clear. If bouncing occurs, then \emph{it seems that the entire collisional process} [...] \emph{is similar to that of elastic balls, which includes a pre-collision phsse, a compression phase, a restitution phase and a post-collision phase} \cite{ShenNature}, and a pile-up of both magnetic flux and plasma may occur. Merging becomes possible when we allow dissipation, which is localized at the reconnection region between plasmoids. As the merging goes on, plasmoids may squeeze themselves on each other like balls of jelly, expanding themselves in the directions $ \perp \textbf{z} $. Then, a fraction of $ E_{K} $ goes into this motion $ \perp \textbf{z} $ during the merging. Accordingly, the absolute value of the inflow speed of plasma into the micro-scale region cannot exceed the velocity of the center of mass of a plasmoid, i.e. $ v_{in} \leq v_{C} $. Now, $ M_{A} = \frac{R}{c_{A} \Delta t} $ and $ \Delta t = \frac{L}{v_{C}} $ lead to $ \frac{v_{C}}{c_{A}} = M_{A} \frac{L}{R} $. Relationships $ v_{in} \leq v_{C} $, $ v_{in} = \frac{R}{\Delta t}$ and $ \Delta t = \frac{L}{v_{C}} $ give $ R \leq L $; then, $ \frac{v_{C}}{c_{A}} = M_{A} \frac{L}{R} $ gives:

\begin{equation} \label{rec01}
v_{C} \geq v_{thr} \quad ; \quad v_{thr} = M_{A} \cdot c_{A}
\end{equation}

(Admittedly, \eqref{rec01} follows trivially from $ v_{in} = \frac{\vert \textbf{E}\vert }{\vert \textbf{B} \vert}$ and $ v_{in} \leq v_{C} $; here we have put in evidence its connection with $ R \leq L $). 
If \eqref{rec01} is violated then the plasmoids approach too slowly to each other, the magnitude of the time derivative of magnetic flux is too low, the induced electric field is too low and Joule dissipation is too weak. Alternatively: if \eqref{rec01} is violated then $ R \geq L $, too many magnetic field lines meet each other outside the plasmoids -  i.e. where there is no plasma and no dissipation - and the braking becomes less effective. Both ways, equation \ref{Econs} remains approximately valid, and leads to bouncing as discussed in Sec. \ref{SEC5} . All the way around, occurrence of bouncing implies negligible dissipation, hence conservation of magnetic flux. Given $ v_{C} $, \eqref{OHM} and \eqref{FARADAY} make such conservation to require an $ \textbf{E} $ with $ \vert \textbf{E} \vert = v_{C} \vert \textbf{B} \vert $. Here, however, the available $ \textbf{E} $ has $ \vert \textbf{E} \vert = M_{A} c_{A} \vert \textbf{B} \vert$; bouncing requires $ v_{C} \vert \textbf{B} \vert < M_{A} c_{A} \vert \textbf{B} \vert$, i.e. violation of \eqref{rec01}.

Further discussion requires more details about dissipation, i.e. on $ M_{A} $. This is a long-standing puzzle of reconnection physics. For a long time now, observations in a wide variety of settings - including reconnection events in solar flares, geomagnetic substorms and sawtooth crashes in magnetically confined fusion devices \cite{Cassak} \cite{Liu0} \cite{Comisso} - have suggested that

\begin{equation} \label{rec02}
M_{A} = \mbox{const.} 
\end{equation}

where - in quite a broad sense - the constant is $ \approx 0.1 $. In particular, according to \cite{Comisso} this statement is true if the flow is incompressible and the outflow velocity is equal to $ c_{A} $; for a significant exception, see \cite{RHESSI}. Incompressibility (i.e., $ v_{C} < c_{sA} $) is compatible with \eqref{rec01} and \eqref{rec02} for $ M_{A} \approx 0.1 $ and $ c_{A} < c_{sA} $. Equation \eqref{rec02} is notoriously in contrast with well-established Sweet and Parker's and Petschek's models of reconnection, as the latter predict much lower values of $ M_{A} $ which, moreover, are decreasing functions of Lundquist number $ S $. As for numerical simulations, fully kinetic computations only seem to be able to reproduce \eqref{rec02} to date \cite{Stanier} \cite{Karimabadi}. However, developing a theoretical understanding of \eqref{rec02} is a challenging task, which lies outside the aim of the present work. For the purpose of our discussion, we are allowed to take \eqref{rec02} as a matter of fact.  Appendix \ref{QUAL3} displays a comparison of \eqref{rec01} and \eqref{rec02} with observations and numerical simulations in both bouncing and merging \cite{ShenNature} \cite{Mishra} \cite{TURNON} \cite{Karimabadi} \cite{Tajima} \cite{Karlicky} \cite{Jelinek}  \cite{RHESSI} \cite{Temmer} . In spite of the huge uncertainties - possibly due to both large measurement errors \cite{Mishra}, insufficient computational resources for realistic problems, and far-from-complete understading of magnetic reconnection - this comparison seems to confirm our result, qualitatively at least. Finally, we discuss a possible connection among  \eqref{variational}, \eqref{rec02} and the results of \cite{Jelinek}, \cite{Liu0}, \cite{Comisso} and \cite{Hesse} in Appendix \ref{QUAL2} .

\section{Conclusions}
\label{SEC6} 

Recently, it has been reported \cite{ShenNature} that collisions between large, magnetized plasmoids produced by coronal mass ejections across the heliosphere are super-elastic, i.e. the linear kinetic energy after the collision is larger than the same energy before the collision. This behavior is similar to what has been observed in collisions of spheres with elastoplastic plates \cite{Louge}. In both cases, increase in kinetic energy occurs at the expense of the energy stored in an internal degree of freedom: in this case, the magnetic field.  

Generally speaking, the plasmoids involved in the scattering are born near the Sun, when solar eruptions occur. In contrast, the scattering itself may occur quite far from the Sun. In the following, by 'far from (near) the Sun' we mean 'at a distance of 1 A.U. (1.8 solar radii) from the center of the Sun'. Available data on temperatures, particle densities and magnetic fields in plasmoids both near and far from the Sun \cite{ShenNature} \cite{TURNON} \cite{ZHOU} \cite{COULD} suggest that both the Lundquist number $ S $ and the Hartmann number $ Ha $ are $ \gg 1 $ everywhere from the very beginning of the flight of plasmoids from the Sun up to the region of space where the scattering occurs. This fact suggests that both viscous and Joule dissipation affect the pattern of streamlines and of magnetic field lines across a plasmoid only weakly during its flight. Then, pattern-related quantities like the magnetic helicity and the generalized helicity discussed in Hall MHD \cite{Witalis} \cite{Turner} behave as global invariant quantities during the flight. (Remarkably, this argument does not require that the evolution of the plasmoid during the flight is self-similar). This is not to say that no dissipation occurs: it may as well raise the internal energy at the expense of mechanical and magnetic energy; but helicities decay much more slowly than energies, and may taken as invariant during the flight. Accordingly, the values of these quantities depend on the structure of the plasmoid near the Sun and may act as suitable initial conditions for the scattering. 

Near the Sun, we show that the evolution of one plasmoid is a succession of quasi-steady states. Analysis of various heating mechanism inside one low-$ \beta $, strongly magnetized (ion Hall parameter $ \Lambda_{i} \gg 1 $) plasmoid shows that each steady state corresponds to a minimum of the total dissipated power \cite{Jaynes}, with the constraints of mass and momentum conservation. The analysis takes advantage of the fact that the evolution of the plasmoid near the Sun is self-similar \cite{Cargill} \cite{ZHOU}. Furthermore, if the interaction of the plasmoid with the external world is  relatively weak in comparison with the magnetic interaction among currents internal to the plasmoid, then previous analytical work \cite{Di Vita00} shows that the solutions of the Euler-Lagrange equations describing this minimum of total dissipated power solve also the Euler-Lagrange equations of a variational principle of Hall MHD. According to the latter principle, each steady state corresponds to a minimum of the sum of kinetic and magnetic energy, with the constraints of fixed magnetic helicity and generalized helicity \cite{Turner}.

Admittedly, the analysis of \cite{Di Vita00} is focussed on plasmoids in the lab. However, the similarities between plasmoids in the lab and in space \cite{Bostick} confirm that quantities - like the generalized helicity - which are relevant to Hall MHD play a role in both lab and space. Indeed, we retrieve the results of \cite{Kagan} and \cite{Ohsaki}: the solar eruption which gives birth to the ejected plasmoid appears to be an instability of a twisted flux rope, in agreement with recent results \cite{Amari}. As a result, a transition occurs between configurations of minimum kinetic + magnetic energy \cite{Turner} endowed with different, Double-Beltrami-like topologies \cite{Kagan}. Transition occurs when the amount of magnetic helicity stored in the system exceeds a threshold (intuitively, when the magnetic field lines are too twisted). The same result has been independently obtained also in \cite{Di Vita01}, where it has been shown that the threshold in helicity is overcome whenever the Hartmann number of the plasmoid exceeds a critical value, a property shared with various coherent magnetized plasma structures in the laboratory \cite{Di Vita00} \cite{Escande} \cite{Bernard} in agreement with an old suggestion of \cite{Bostick1}.

Above threshold, the instantaneous configuration of one plasmoid is well-described with the help of Taylor’s variational principle \cite{Taylor} which minimizes the magnetic energy with the constraint of fixed magnetic helicity, a principle whose Euler-Lagrange  equation for the magnetic field describes force-free equilibria often invoked in plasma physics both in laboratory \cite{Bodin} \cite{Bellan} and space \cite{Low} \cite{Choudhuri}. The resulting magnetic field of the spheromak is force-free \cite{Taylor} and spheromak-like \cite{Bellan}; thus, we retrieve as a consequence of our discussion the results postulated without proof in \cite{ZHOU} and \cite{WANG}.

After a flight started in the neighbourhood of the Sun, two plasmoids (say, 1 and 2) bump into each other, each one with its own magnetic helicity and generalized helicity inherited from the birth near the Sun. As for their scattering, it seems to be quite a short event. More precisely, it is so short that diffusion of particles and of magnetic field lines is not in time to affect the values of total magnetic and generalized helicity. For simplicity, we discuss head-on collision along an axis $ \textbf{z} $ of two plasmoids with the same mass, in the frame of reference of the center of mass. Four  conservation equations - concerning total mass, total magnetic helicity, total generalized helicity and total energy - correspond to the four relevant conditions of conservation of total mass, energy, magnetic flux and the component of momentum along $ \textbf{z} $. 

We investigate the behaviour of the system under a scaling transformation. The latter leaves both mass, energy and twistedness of streamlines and magnetic field lines (hence the magnetic and the generalized helicities) unaffected. Starting from this fact, we show that any physical process (no matter what its underlying mechanism is like) leading to an enlargement of a plasmoid raises its linear kinetic energy at the expense of the magnetic energy, as suggested in \cite{ShenNature}. This result relies on no assumption concerning the detailed description of the processes occuring inside the plasmoids during the scattering. 

In particular, the approach of plasmoid 2 to plasmoid 1 changes the magnetic flux across 1 and drives Faraday currents across it (and vice versa). The resulting dissipation heats 1; as a consequence, it makes 1 to expand and - according to our result above - raises its linear kinetic energy. Generally speaking, the larger the relative growth of the latter quantity in comparison with its initial value, the lower the impact velocity, just as suggested in \cite{ShenNature}.

So far, we have discussed the ('bouncing') case where the plasmoids move towards (away from) each other before (after) the collision \cite{ShenNature} \cite{Mishra} \cite{Karimabadi}. However, the nature of collision between plasmoids have been investigated by various authors and is found that its regime can range from super-elastic to inelastic \cite{Mishra}. Two colliding plasmoids may just merge \cite{Karimabadi} \cite{Tajima} \cite{Karlicky} \cite{Jelinek} \cite{RHESSI}.

Intuitively, we expect a large change in magnetic topology in case of merging, so that the occurrence of the latter depends on magnetic reconnection. Even if our understanding of reconnection in this large-$ S $ problem is still far from complete, a simple model (in analogy with elastoplastic collisions \cite{Magnanimo}) suggests that the choice between bouncing and merging depends on the relative velocity of the plasmoids. Indeed, the larger such velocity, the faster the change of magnetic flux, the larger the induced electric field and the corresponding dissipation, which is related to modifications of magnetic topology. The model does not depend on detailed knowledge of ion dynamics (which defies current fluid models \cite{Stanier} \cite{Ng}), and leads to a simple rule-of-thumb: if the impact velocity is greater (less) than the product of Alfvèn speed times the global reconnection rate then merging (bouncing) occurs. Above threshold, dissipation slows down the relative motion of plasmoids until merging occurs. Superelastic scattering occurs in the extreme opposite case where the impact velocity is much less than this threshold. 

In spite of the uncertainties - due to both large measurement errors, insufficient computational resources for realistic problems, and far-from-complete understading of magnetic reconnection - comparison with observations \cite{ShenNature} \cite{RHESSI} and numerical simulations \cite{Stanier} \cite{Karimabadi} \cite{Tajima} \cite{Karlicky} \cite{Jelinek} seem to confirm our result, qualitatively at least. In all these cases, we follow \cite{Cassak} \cite{Liu0} \cite{Comisso} and assume the value of the global reconnection rate to be $ \approx 0.1 $. This value does not depend on $ S $, in contrast with the predictions of Sweet and Parker's and Petschek's models of reconnection. (The model takes no shock wave into account. The latter may raise dissipation and make therefore the requirement on impact velocity for merging harder to satisfy. This agrees qualitatively with observations \cite{Mishra}). 
 
We may justify this assumption as follows. As merging goes on, dissipation lowers both magnetic and kinetic energy much more rapidly than the magnetic and kinetic helicity, so that the values of the former (the latter) drop further quickly (change slowly). We are therefore allowed to describe the evolution of the system as a succession of relaxed, steady states - each of them satisfying the variational principle discussed above - provided that the relaxation time (i.e. the time-of-flight of magnetosonic waves across the plasmoids) is short enough. Here the word 'steady' has a statistical meaning only, as 
the complexity of the dynamics gives rise to unsteady processes at large $ S $, which cannot be described by Sweet and Parker's and Petschek's steady-state models.

A relaxed state enjoys the following properties. Firstly, and regardless of the detailed ion dynamics, electrons follow the path of lesser resistance, just like in the electrical networks on Earth. Correspondingly, the global reconnection rate is a maximum. Near this maximum, it depends very weakly on the detailed geometry of field lines. The latter is affected by the motion of the merging plasmoids. As a consequence, once the reconnection rate has achieved a maximum it sticks pretty much to the same value as the merging goes on, i.e. it remains constant as the merging goes on. Finally, the breadth of the maximum of allows the same value of the reconnection rate to apply to a wide range of disparate (even relativistic) physical systems.  

In a nutshell: taking advantage of concepts originally developed for plasmoids in the lab \cite{Bostick} \cite{Bostick1} \cite{Witalis} \cite{Turner} \cite{Auluck1} \cite{Auluck2} \cite{Di Vita02}, we have shown that the super-elastic nature of plasmoid-plasmoid scattering at low impact velocity \cite{ShenNature} follows from both the invariance of the scattering physics (no matter how complex it may be) under rescaling and from: 

a) the conservation of global invariant quantities (magnetic and generalized helicity) at large values of $ S $ and $ Ha $, both near and far from the Sun; 

b) near the Sun, the modeling of the solar eruption which gives birth to each plasmoid as an instability of a twisted flux rope \cite{Amari}, where the evolution of the latter is described as a succession of relaxed, Double-Beltrami plasma structures \cite{Kagan} \cite{Ohsaki} and the outgoing plasmoid \cite{ZHOU} is a force-free \cite{WANG}, spheromak-like \cite{Bellan} structure satisfying Taylor's principle \cite{Taylor}.  

Remarkably, a) is a tenet of Hall MHD \cite{Witalis} \cite{Turner}. It links what happens at the scattering far from the Sun and what has occurred at the birth of each plasmoid, near the Sun. 
Moreover, and in contrast with previous literature, b) is no assumption. 
Rather, it is a consequence of the identification of the relaxed structures with minima of the dissipated (viscous + Joule) power. 
In turn, we have invoked the results of \cite{Di Vita00}, generalized the results of \cite{Lamb}, \cite{Jaynes}, \cite{PRE2010}, \cite{Dasgupta} and \cite{Dasgupta2} and shown that the latter identification follows from the structure of Ohm's law and of the viscous stress tensor in our low-$ \beta $, $ \Lambda \gg 1 $, weakly dissipative plasma near the Sun. The proof holds for vanishing temperature gradient only, but can easily be generalized to the case of self-similar evolution \cite{Cargill} \cite{ZHOU}.

The larger the impact velocity, the less evident the superelastic nature of the scattering. Above a threshold - approximately, Alfvèn velocity times the global reconnection rate - the scattering becomes inelastic altogether, and the colliding plasmoids may merge \cite{Tajima} \cite{Karlicky} \cite{Karimabadi} \cite{Jelinek} \cite{RHESSI}. As the merging goes on, electrons follow the path of lesser resistance, the value of the global reconnection rate remains approximately constant, and near to the value observed in a wide variety of settings \cite{Cassak} \cite{Liu0} \cite{Comisso}. Observations \cite{ShenNature} \cite{RHESSI} and simulations \cite{Stanier} \cite{Karimabadi} \cite{Tajima} \cite{Karlicky} \cite{Jelinek} confirm our result.

Admittedly, in the present treatment we obtain no more than a qualitative agreement with observations. For example, we grossly overestimate the enlargement of plasmoids at the end of the superelastic scattering; moreover, the estimate of the global reconnection rate is  approximate. In our discussion, however, and in contrast with previous analysis \cite{Cargill}, the role of geometrical quantities (the helicities) and dynamical quantities (mass, energy) are dealt with on an equal footing. Further analysis requires thorough, detailed balance of energy, mass etc. during the scattering. This will be the topic of future work.

\section*{Acknowledgments}

Useful discussions and warm encouragement with Prof. W. Pecorella, Università di Tor Vergata, Roma, Italy are gratefully acknowledged.

\section*{Author contribution statement}

A. Di Vita is the one and only author of this paper.

\appendix 

\section{Proof of \eqref{LJAPUNOV}}
\label{QUAL}

According to \eqref{PV} and \eqref{PJoule}, in the $ \Lambda_{i} \gg 1 $ limit 
$ \int_{\Omega} P_{h} d^3\mbox{x} $ depends on both $ \textbf{v} $, $ \textbf{B} $, 
$ \bm{\eta}_{\left( V \right)} $ and $ \bm{\eta}_{\left( J \right)} $. Perturbations 
of $ \textbf{v} $ leave both $ \bm{\eta}_{\left( V \right)} $ and $ \bm{\eta}_{\left( J \right)} $ unaffected. The effect of perturbations of 
$ \textbf{B} = \textbf{b} \vert \textbf{B} \vert $ on $ \bm{\eta}_{V} $ and 
$ \bm{\eta}_{J} $ is the sum of the result of 
the perturbation of the unit vector $ \textbf{b} $ and of $ \vert \textbf{B} \vert $. Any 
perturbation of $ \textbf{b} $ leaves its squared norm $ \textbf{b} \cdot \textbf{b} = 1$ 
unaffected; such unitary transformation, a rotation in real space, leaves true scalars 
like $ P_{v} $, $ P_{J} $ and $ P_{h} $ unaffected. As for the perturbations of 
$ \vert \textbf{B} \vert $, no component of $ \bm{\eta}_{\left( J \right)} $ depends on this quantity, 
and the components of $ \bm{\eta}_{\left( V \right)} $ depend on it only through powers of 
$ \Lambda_{i}^{-1} $; derivatives of the latter components on $ \vert \textbf{B} \vert $ 
are therefore $ \propto $ powers of $ \Lambda_{i}^{-1} $ and are therefore negligible in 
the $ \Lambda_{i} \gg 1 $ limit. After repeated integration by parts with vanishing $ \frac{\partial \textbf{v}}{\partial t} $ and $ \frac{\partial \textbf{B}}{\partial t} $ on the boundary, equations \eqref{NS}-\eqref{PJoule} lead therefore to:

\begin{eqnarray}
\begin{split}
&
\dfrac{\partial}{\partial t} \left( \int_{\Omega} P_{h} d^3\mbox{x} \right) = 
2 \int_{\Omega} \textbf{j} \cdot \bm{\eta_{\left( J \right)}} \cdot \dfrac{\partial \textbf{j}}{\partial t} d^3\mbox{x} +
2 \int_{\Omega} 
\eta_{\left( V \right) ijkl} \dfrac{\partial v_{k}}{\partial x_{l}} 
\dfrac{\partial}{\partial t} 
\left( \dfrac{\partial v_{i}}{\partial x_{j}} \right) 
d^3\mbox{x} 
+ O \left( \Lambda_{i}^{-1} \right) 
=
\\
&
= 
2 \int_{\Omega} 
\textbf{j} 
\cdot 
\dfrac{\bm{\eta_{\left( J \right)}}}{\mu_{0}}
\cdot 
\nabla \wedge \dfrac{\partial \textbf{B}}{\partial t} 
d^3\mbox{x} 
+ 
2 \int_{\Omega} 
\tau_{ij} 
\dfrac{\partial}{\partial x_{j}} 
\left( \dfrac{\partial v_{i}}{\partial t} \right)
d^3\mbox{x} 
+ O \left( \Lambda_{i}^{-1} \right) 
=
\\
&
= 
2 \int_{\Omega} 
\dfrac{1}{\mu_{0}}
\dfrac{\partial \textbf{B}}{\partial t}
\cdot 
\nabla 
\wedge
\left( \textbf{j} \cdot \bm{\eta_{\left( J \right)}} \right) 
d^3\mbox{x} 
- 2 \int_{\Omega} 
\dfrac{\partial \textbf{v}}{\partial t} 
\cdot
\left( \nabla \cdot \bm{\tau} \right) 
d^3\mbox{x} 
+ O \left( \Lambda_{i}^{-1} \right) 
=
\\
&
= 
2 \int_{\Omega} 
\dfrac{1}{\mu_{0}}
\dfrac{\partial \textbf{B}}{\partial t}
\cdot 
\nabla 
\wedge
\left( \textbf{E} + \textbf{v} \wedge \textbf{B} \right) 
d^3\mbox{x} 
+ O \left( \beta \right) 
+ O \left( \left( \vert \textbf{k} \vert L \right)^{-1} \right) 
+
\\
&
- 2 \int_{\Omega} 
\rho
\vert \dfrac{\partial \textbf{v}}{\partial t} \vert^2 
d^3\mbox{x} 
- 2 \int_{\Omega} 
\rho
\dfrac{\partial \textbf{v}}{\partial t} 
\cdot
\left[ \left( \textbf{v} \cdot \nabla \right) \textbf{v} \right]
d^3\mbox{x} 
- 2 \int_{\Omega} 
\dfrac{\partial \textbf{v}}{\partial t} 
\cdot
\nabla \left( p + \rho \phi_{g} \right)
d^3\mbox{x} 
+ O \left( \Lambda_{i}^{-1} \right) 
=
\\
&
= 
- 
2 \int_{\Omega} 
\dfrac{1}{\mu_{0}}
\vert
\dfrac{\partial \textbf{B}}{\partial t}
\vert^2 
d^3\mbox{x} 
+ 
2 \int_{\Omega} 
\dfrac{1}{\mu_{0}}
\dfrac{\partial \textbf{B}}{\partial t}
\cdot 
\nabla 
\wedge
\left( \textbf{v} \wedge \textbf{B} \right) 
d^3\mbox{x} 
+ O \left( \beta \right) 
+ O \left( \left( \vert \textbf{k} \vert L \right)^{-1} \right) 
+ 
\\
&
- 2 \int_{\Omega} 
\rho
\vert \dfrac{\partial \textbf{v}}{\partial t} \vert^2 
d^3\mbox{x} 
+ 2 \int_{\Omega} 
\rho
\dfrac{\partial \textbf{v}}{\partial t} 
\cdot
\left[ \textbf{v} \wedge \left( \nabla \wedge \textbf{v} \right) \right]
d^3\mbox{x} 
- 2 \int_{\Omega} 
\dfrac{\partial \textbf{v}}{\partial t} 
\cdot
\nabla \left( p + \rho \phi_{g} + \rho \dfrac{\vert \textbf{v} \vert^2}{2}\right)
d^3\mbox{x} 
+ O \left( \Lambda_{i}^{-1} \right) 
=
\\
&
= 
- 
2 \int_{\Omega} 
\dfrac{1}{\mu_{0}}
\vert
\dfrac{\partial \textbf{B}}{\partial t}
\vert^2 
d^3\mbox{x} 
- 2 \int_{\Omega} 
\rho
\vert \dfrac{\partial \textbf{v}}{\partial t} \vert^2 
d^3\mbox{x} + O \left( \beta \right) + O \left( \left( \vert \textbf{k} \vert L \right)^{-1} \right) + O \left( \Lambda_{i}^{-1} \right) + 
\\
&
+ 
2 \int_{\Omega} 
\dfrac{1}{\mu_{0}}
\dfrac{\partial \textbf{B}}{\partial t}
\cdot 
\nabla 
\wedge
\left( \textbf{v} \wedge \textbf{B} \right) 
d^3\mbox{x} 
+
2 \int_{\Omega} 
\rho
\dfrac{\partial \textbf{v}}{\partial t} 
\cdot
\left[ \textbf{v} \wedge \left( \nabla \wedge \textbf{v} \right) \right]
d^3\mbox{x}
\\
\end{split}
\end{eqnarray}

where we have taken into account that $ \nabla p_{e} \propto \nabla p, \textbf{j} \wedge \textbf{B} \approx \nabla p \approx O \left( \beta \right), \textbf{R}_{T} \propto O \left( \nabla T \right) $ and that $ T = T_{b} $. Inequality \eqref{LJAPUNOV} follows for a $ \vert \textbf{k} \vert L \gg 1 $ perturbation of $ \textbf{v} $ and $ \textbf{B} $ in a $ \beta \ll 1, \Lambda_{i} \gg 1$ plasmoid if 
$ \int_{\Omega} 
\frac{1}{\mu_{0}}
\frac{\partial \textbf{B}}{\partial t}
\cdot 
\nabla 
\wedge
\left( \textbf{v} \wedge \textbf{B} \right) 
d^3\mbox{x} 
$ 
and 
$
\int_{\Omega} 
\rho
\frac{\partial \textbf{v}}{\partial t} 
\cdot
\left[ \textbf{v} \wedge \left( \nabla \wedge \textbf{v} \right) \right]
d^3\mbox{x} $ 
are negligible. This is true for the former (the latter) quantity e.g. if $ Re \ll 1 $ ($ Re_{m} \ll 1$), as in the original proof of Kortweweg-Helmholtz' (Kirchhoff's) principle. 
Here, however, this is not true. We are going to show that both $ \int_{\Omega} 
\frac{1}{\mu_{0}}
\frac{\partial \textbf{B}}{\partial t}
\cdot 
\nabla 
\wedge
\left( \textbf{v} \wedge \textbf{B} \right) 
d^3\mbox{x} 
$ 
and 
$
\int_{\Omega} 
\rho
\frac{\partial \textbf{v}}{\partial t} 
\cdot
\left[ \textbf{v} \wedge \left( \nabla \wedge \textbf{v} \right) \right]
d^3\mbox{x} $ 
are negligible if $ \beta \ll 1, Re_{m} \gg 1, \vert \textbf{k} \vert L \gg 1 $ and $ \nabla T = 0 $.

As for $ \int_{\Omega} 
\frac{1}{\mu_{0}}
\frac{\partial \textbf{B}}{\partial t}
\cdot 
\nabla 
\wedge
\left( \textbf{v} \wedge \textbf{B} \right) 
d^3\mbox{x} 
$ , integration by parts shows that  
$ \int_{\Omega} \frac{1}{\mu_{0}}
\frac{\partial \textbf{B}}{\partial t}
\cdot 
\nabla 
\wedge
\left( \textbf{v} \wedge \textbf{B} \right) 
d^3\mbox{x} =
\int_{\Omega} 
\frac{\partial \textbf{j}}{\partial t}
\cdot 
\left( \textbf{v} \wedge \textbf{B} \right) 
d^3\mbox{x} $; since $ \textbf{j} \wedge \textbf{B} \approx O \left( \beta \right) $, we write $ \textbf{j} = \alpha \textbf{B} + O \left( \beta \right), \alpha \equiv \frac{\textbf{j} \cdot \textbf{B}}{\vert \textbf{B} \vert^2} $. Then, $ \frac{\partial \textbf{j}}{\partial t}
\cdot 
\left( \textbf{v} \wedge \textbf{B} \right) = 
\frac{\partial \alpha}{\partial t} \textbf{B} \cdot 
\left( \textbf{v} \wedge \textbf{B} \right) +
\alpha 
\frac{\partial \textbf{B}}{\partial t} \cdot \left( \textbf{v} \wedge \textbf{B} \right) 
+ O \left( \beta \right) = \alpha 
\textbf{v} \cdot \left( \textbf{B} \wedge \frac{\partial \textbf{B}}{\partial t} \right) 
+ O \left( \beta \right) = 
\frac{\alpha}{2} \frac{\partial}{\partial t} \left( \textbf{B} \wedge \textbf{B} \right) 
+ O \left( \beta \right) = O \left( \beta \right)$. 

As for  
$
\int_{\Omega} 
\rho
\frac{\partial \textbf{v}}{\partial t} 
\cdot
\left[ \textbf{v} \wedge \left( \nabla \wedge \textbf{v} \right) \right]
d^3\mbox{x} $, equation \eqref{OHM} gives $ \vert \textbf{v} \vert \approx \frac{ \vert \textbf{E} \wedge \textbf{B} \vert}{\vert \textbf{B} \vert^2} + O \left( \beta \right) + O \left( \left( \vert \textbf{k} \vert L \right)^{-1} \right) + O \left( Re_{m}^{-1} \right) $. Moreover, \eqref{FARADAY} ensures that $ \vert \textbf{E} \vert \propto \vert \frac{\partial \textbf{B}}{\partial t} \vert$, so that 
$ \frac{\partial \textbf{v}}{\partial t} 
\cdot
\left[ \textbf{v} \wedge \left( \nabla \wedge \textbf{v} \right) \right]
\approx 
O \left( \vert \frac{\partial^2 \textbf{B}}{\partial t^2} \vert \vert \frac{\partial \textbf{B}}{\partial t} \vert^2 \right) 
+ O \left( \beta \right) + O \left( \left( \vert \textbf{k} \vert L \right)^{-1} \right) + O \left( Re_{m}^{-1} \right)$. Unless the decay of the relaxation is quite fast, we may safely neglect $ O \left( \vert \frac{\partial^2 \textbf{B}}{\partial t^2} \vert \vert \frac{\partial \textbf{B}}{\partial t} \vert^2 \right) $: this is the meaning of the word 'slow' in the text. Under our assumptions, all other terms are negligible, and \eqref{LJAPUNOV} follows.

\section{Comparison of \eqref{rec01} and \eqref{rec02} with available data}
\label{QUAL3}

As for observations of bouncing, two plasmoids (1 and 2) have been observed in \cite{ShenNature}. Being faster than 1, 2 finally caught up and collided with 1; super-elastic bouncing followed. According to Table 1 of \cite{ShenNature}, the components $ v_{1} $ and $ v_{2} $ along the direction of collision of the velocities of plasmoid 1 and 2 moving away from the Sun are 205 Km/s and 237 Km/s respectively in the heliocentric frame of reference. Thus, the absolute value $ v_{r} \equiv \vert v_{2} - v_{1} \vert $ of the component of the relative velocity along the same direction is equal to 237 - 205 = 32 Km/s. The masses $ m_{1} $ and $ m_{2} $ of 1 and 2 being of the same order of magnitude, we write $ \vert v_{C} \vert \approx \frac{v_{r}}{2} = 16 $ Km/s. For comparison, $ c_{A} \approx 100$ Km/s inside the plasmoids, so that $ \frac{v_{C}}{c_{A}} \approx 0.16 $. 

As for simulations of bouncing, we refer to one of the cases described in \cite{Karimabadi}. Along the direction of collision, the travel $ \Delta L $ of a plasmoid before collision is $ \approx 75 \cdot \frac{c}{\omega_{pi}} $ long. The duration $ \Delta t $ of the event is about 3 times the Alfvèn time $ \frac{\Lambda}{c_{A}} $, where $ \Lambda = \alpha \cdot \frac{c}{\omega_{pi}}$ is a typical linear size and $ \alpha $ may take different values. Then, $ \frac{v_{C}}{c_{A}} = \frac{\Delta L}{c_{A} \cdot \Delta t}  \approx \frac{25}{\alpha} $. If $ \alpha = 100 $ then $ \frac{v_{C}}{c_{A}} = 0.25 $ and bouncing occurs, in qualitative agreement with \eqref{rec01}. 

As for observations of merging, we observe that our condition \eqref{rec01} for merging holds regardless both of the relative size of the two plasmoids and of the role of gravity. Thus, we try to apply it also to the observed merging of a plasmoid with a loop-top kernel near the Sun \cite{RHESSI}. The mass of the plasmoid being $ \ll $ the mass of the target loop-top kernel, we may safely identify $ v_{C} $ with the observed value of plasmoid velocity, $ \approx 12 $ Km/s; for comparison, $ c_{A} \approx 400 $ Km/s and (in this case) $ M_{A} = 0.001 $, so that \eqref{rec01} is satisfied. 
This result clearly shows that it is the value of $ \frac{v_{C}}{M_{A} \cdot c_{A}} $ which is relevant, rather than the values of $ \frac{v_{C}}{c_{A}} $ and $ M_{A} $ separately.

As for simulations of merging, in the $\left( 2 + \frac{1}{2} \right)-$D computation of \cite{Tajima} the velocity $ v_{C} $ in a system with typical length $ \Delta L = 128 \cdot \lambda_{D} $ where the duration of the travel of a plasmoid is $ \Delta t = \frac{24}{\Omega_{e pol}}$ is $ v_{C} = \frac{\Delta L}{\Delta t} $, where $ \Omega_{e pol} $, $ \lambda_{D} = \frac{v_{the}}{\omega_{pe}}$ and $ v_{the} $ are the electron cyclotron frequency in the poloidal component of $ \textbf{B} $, the Debye length and the electron thermal speed respectively. In turn, $ \Omega_{e pol} = 0.77 \cdot \omega_{pe} $, so that  $ v_{C} = 6.93 \cdot v_{the} $. Moreover, the electron cyclotron frequency in the toroidal component of $ \textbf{B} $ is $ \Omega_{e tor} = 0.2 \cdot \omega_{pe} $, and the Alfvèn velocity in the poloidal component of $ \textbf{B} $ is $ c_{A pol} = 1.22 \cdot v_{the} $, where $ c_{A} = c_{A pol} \cdot \sqrt{1 + \frac{\Omega_{e tor}}{\Omega_{e pol}}} $. It follows that $ v_{the} = 0.73 c_{A} $, so that $ \frac{v_{C}}{c_{A}} = 5 $; accordingly, \eqref{rec01} and \eqref{rec02} predict occurrence of merging, in agreement with the results of \cite{Tajima}.

Ref. \cite{Karlicky} provides us with a further example of simulation of plasmoid merging. Here $ \lambda_{D} = 0.6 \cdot \Delta $ and $ \frac{c}{\omega_{pe}} = 10 \cdot \Delta $ where $ \Delta $ is the spatial step of the simulation is $ \Delta $. Accordingly, $ \frac{v_{the}}{c} = \frac{\lambda_{D} \cdot \omega_{pe}}{c} = 0.06 \cdot c $. The thermal speed of ions is $ v_{thi} = v_{the} \cdot \sqrt{\frac{m_{i}}{m_{e}}} $, and the value of $ \beta = 0.07 $ leads to $ c_{A} = \sqrt{\beta} \cdot v_{thi} \approx 100$ Km/s. Merging occurs on a spatial scale $ \Delta L = 3000 \cdot \Delta \left( = 5000 \cdot \lambda_{D} \right)$ in a time-scale $ \Delta t = \frac{6500}{\omega_{pe}} $, hence $ v_{C} = \frac{\Delta L}{\Delta t} = 0.77 \cdot v_{the} \approx 0.05 \cdot c \approx 150 \cdot c_{A}$, i.e. $ \frac{v_{C}}{c_{A}} \approx 150 $ and merging is predicted by \eqref{rec01} and \eqref{rec02}, in agreement with the results of \cite{Karlicky}.

In the simulations of Ref. \cite{Jelinek}, $ c_{A} = 2.15 \cdot 10^5 $ m/s. Moreover, Fig. 2 and 3 display two similar plasmoids approaching to each other. Their distance decreases by $ 5 \cdot 10^6 $ m in 10 s, i.e. $ v_{r} = \frac{5 \cdot 10^6}{10} = 5 \cdot 10^5 $ m/s. Taking as usual $ \vert v_{C} \vert $ to be half such value, we obtain $ v_{C} = 2.5 \cdot 10^5  $ m/s, i.e. $ \frac{v_{C}}{c_{A}} \approx 1 $ and merging is predicted by \eqref{rec01} and \eqref{rec02}, in agreement with the results of \cite{Jelinek}.

Furthermore, if we take $ \alpha \leq 25 $ in the simulations of Ref. \cite{Karimabadi} then $ \frac{v_{C}}{c_{A}} \geq 1 $ and merging is predicted by \eqref{rec01} and \eqref{rec02}, in agreement with the results of \cite{Karimabadi}. The fact that the larger the value of $ \alpha $ the easier the bouncing underlines the relevance of ion motion \cite{Comisso}, which fluid models may fail to grasp (thus overestimating the likelihood of merging) \cite{Stanier}.

Our discussion takes into account no shock wave. Indeed, observations suggest that plasmoid 2 did not drive an evident shock ahead in \cite{ShenNature}: however, it has been observed in \cite{TURNON} that the approaching speed in \cite{ShenNature} is relatively low, in comparison with other observations \cite{Mishra} \cite{Temmer} at least. Shock waves are dissipative phenomena; if they occur, they make the system more dissipative and we expect the condition \eqref{rec01} for merging to become more stringent as the supply of kinetic energy linked with $ \vert v_{C} \vert $ and required to sustain dissipation increases if shocks are added. Since dissipation is taken into account in the R.H.S. of \eqref{rec01}, this is equivalent to say that we expect \eqref{rec01} still to hold when shock waves are present, but with a somewhat larger value of $ v_{thr} $: if shocks are present, then merging is more difficult. Indeed, the value of $ M_{A} $ may strongly differ from $ 0.1 $ if compressibility plays a role \cite{Comisso}.

An observation of bouncing plasmoids is reported in Sec. 3.2 of \cite{Mishra}, which is compatible with the occurrence of shock waves. Again, 2 crashes into 1 in a rear-end collision; reportedly, the initial speeds and the directions of flight of 1 and 2 are 385 km/s, 19°W, 11°S and 610 km/s, 25°W, 13°N respectively, so $ v_{r} \approx $ 218 Km/s. Taking as usual $ \vert v_{C} \vert $ to be half such value, we obtain $ v_{C} = 109 $ Km/s. Now, Fig. 7 of \cite{Mishra} provides us with typical values $ \vert \textbf{B} \vert = 20 $ nT, $ n_{i} = 10 $ cm$^{-3}$, hence $ c_{A} \approx 140 $ Km/s. Thus, bouncing with shock waves occurs with $ \frac{v_{C}}{c_{A}} \lesssim 1 $. 

Another observation of bouncing plasmoids which is compatible with shock waves is reported in \cite{Temmer}, where the initially faster plasmoid 2 (slower plasmoid 1) decelerates (accelerates) from $ \approx 1300 $ Km/s to $ \approx 600 $ Km/s (from $ \approx 400 $ Km/s to $ \approx 700 $ Km/s). Both masses and directions of propagation are basically the same. Reasonably, we may take $ v_{C} $ to be of the same order of magnitude of $ c_{A} $, i.e. $ \frac{v_{C}}{c_{A}} \approx 1 $ holds.

Finally, the results of numerical computations of $ v_{thr} = v_{thr} \left( v_{1} \right) $ are reported in \cite{TURNON}, where $ m_{1} = m_{2}$ and both $ v_{2} $ and the ratio $ \kappa $ of kinetic energy and total energy of plasmoid 1 $ \left( 0 < \kappa < 1 \right) $ are kept fixed. If $ \kappa \rightarrow 1 $, 1 is fast, $ v_{1} \rightarrow c_{sA}$ and shock waves are expected to occur both ahead and behind the plasmoid. According to our discussion above, the nearer $ v_{1} $ to the velocity of one of these shock waves, the stronger the dissipation required for inelastic scattering, the larger $ v_{thr} $. Then, we expect the function $ v_{thr} = v_{thr} \left( v_{1} \right) $ to have two peaks corresponding to the velocities of the ahead-shock and behind-shock wave. Numerical simulations seem to confirm this conclusion, qualitatively at least - we refer to the red and green lines in Fig. 4.a of \cite{TURNON}. In contrast, if $ \kappa \rightarrow 0 $ then no shock wave is expected to occur and \eqref{rec01} forbids elastic scattering when $ v_{C} $ exceeds a threshold. All the way around, the lower $ v_{C} = \frac{v_{r}}{2} = \frac{\vert v_{2}-v_{1} \vert}{2}$ the higher the values of $ v_{1} $, of the kinetic energy of plasmoid 1, of its magnetic energy, of $ \vert \textbf{B} \vert $ and of $ \Lambda_{i} $ in 1, the easier the violation of \eqref{thr1}, and the more unlikely the preservation of the Taylor-like structure of the scattered plasmoid, i.e. dissipation rules and the scattering cannot be elastic - all the more so, as \eqref{thr2} is barely satisfied far from the Sun. We conclude that elastic scattering is only possible within a narrow interval of values of $ v_{C} $ - hence of $ v_{1} $. Accordingly, we expect $ v_{thr} = v_{thr} \left( v_{1} \right) $ to differ from zero in a narrow interval of values of $ v_{1} $ only. Again, simulations provide us with qualitative confirmation of this conclusion - see the black and blue lines in Fig. 4.a of \cite{TURNON} .

\section{More about \eqref{rec02}}
\label{QUAL2}

When $ S $ is large enough, the value of $ M_{A} $ tends to be the same in most models. This suggests that the reconnection rate is set not by the micro-scale physics allowing the dissipation, but is due to constraints at large  scales since the only aspect many different forms of reconnection have is that they match up with MHD at large scales \cite{Cassak}. 

Magnetic reconnection is a dissipative phenomenon. The rate of reconnection is likely related to the efficiency of particle acceleration and heating during the reconnection process. Reconnection tends to occur wherever strong currents concentrate, i.e. we expect it to occur within relatively small regions. Thus, the Joule heating due to the reconnection event itself is small due to the small size of dissipation region and a low resistivity. A part of the magnetic energy released is converted into kinetic energy. The heating usually occurs away from the reconnection region proper via a number of non-ideal plasma processes (shocks, waves, adiabatic heating, viscous heating). This implies that heating occurs in a much larger volume than the volume of the micro-scale region. When it comes to write down the required large-scale constraint on dissipative magnetic reconnection, therefore, the total dissipated power $ \int_{\Omega} P_{h} d^3\mbox{x} $ is an obvious candidate. Here we suggest that the relevant constraint (for our problem of merging plasmoids at least) is provided by \eqref{variational}. In the following, we are going to justify this assumption, to draw the consequences and to show that our results fit the results of Ref. \cite{Liu0} and justify \eqref{rec02}. Our discussion is qualitative; then, the operator '$ = $' refers to order-of-magnitude-estimates only in the following.
 
Dissipation leaves both $ K $ and $ H $ unaffected, while raising internal energy and further decreasing $ E $. Thus, the system made of the two merging plasmoids attains ever lower values of $ E $ at constant $ K $ and $ H $, even if the value of $ \beta $ is large enough to invalidate our proof of \eqref{LJAPUNOV} in Appendix \ref{QUAL}. Accordingly, we may still describe the evolution of the system as a succession of steady-state solutions of Turner's variational principle \cite{Turner}, which are just (possibly non-Taylor) DB steady states. This assumption makes sense provided that, as usual by now, we neglect both $ \nabla T $ and the relaxation time-scale $ \tau $. In turn, the latter assumption is reasonable  e.g. if the relaxation time-scale is $ \tau = \frac{L}{c_{sA}} $ and if $ \tau < \Delta t = \frac{L}{v_{C}}$, i.e. $ v_{C} < c_{sA} $; in turn, this is compatible with \eqref{rec01} if $ M_{A} < 1 $ as $ c_{sA} > c_{A} $. This argument agress with \cite{Jelinek}: merging plasmoids undergo oscillations, which are associated with magnetoacoustic waves produced by the motion and merging of plasmoids; these oscillations are quickly damped by the plasma flows in the vicinity of the oscillating plasmoid.

Admittedly, we have not yet dealt with the thin reconnection layer between the colliding plasmoids. However, this is likely to leave our argument unaffected. Large $ S $ reconnection is unsteady due to the continuous formation, merging, and ejection \cite{Comisso} of tiny plasmoids \cite{Karlicky}. Again, the latter undergo damped magnetosonic oscillations \cite{Jelinek}. Since the reconnection layer is much smaller than the colliding plasmoids which it lies between, the time-of-flight of magnetosonic waves across the reconnection layer is even shorter than $ \tau $; then, the evolution of the reconnection layer as well may be described as a succession of  marginally stable, steady states - where the word 'steady' has just a statistical meaning \cite{Comisso}. (Remarkably, reconnection in electron-positron plasmas seems to follow \eqref{rec01} and \eqref{rec02} with $ M_{A} \approx 0.1 $ \cite{Cassak}; and relaxed states in such plasma are precisely $ \int_{\Omega} P_{h} d^3\mbox{x} = \min. $, DB states \cite{Dasgupta3}). In the following, by 'system' we are going to refer to 'the system made of both the two colliding plasmoids and of the reconnection layer between them'. 

We have seen in Sec. \ref{SEC4} that a DB steady state corresponds also to a solution of \eqref{variational}. The latter includes both viscous and Joule dissipation. Viscous dissipation has been taken into account in the proof \cite{Di Vita01} of \eqref{thr1}. Here we focus on Joule dissipation only, which is ruled by electrons and does not therefore depend on the meandering orbits of ions. 
Accordingly, we focus on Kirchhoff's principle $ \int_{\Omega} P_{J} d^3\mbox{x} = \min $ with the constraint $ \nabla \cdot \textbf{j} = 0 $ of electric charge conservation. In other words, we are going to describe the distribution of electric current density $ \textbf{j} $ across the system at a given time as a relaxed state satisfying Kirchhoff's principle. We want to take advantage of Kirchhoff's principle in order to obtain information about $ M_{A} $. Our strategy is to reformulate it in a way which contains $ M_{A} $ in a simple way. We discuss the constraint and the minimized quantity separately. 

As for $ \nabla \cdot \textbf{j} = 0 $, it allows the value $ I = \frac{\int_{\Omega} P_{J} d^3\mbox{x}}{V} $ of electric current flowing across the surface of area $ \frac{\Delta \Phi}{\vert \textbf{B} \vert} $ crossed by the magnetic field lines during the reconnection (see main text) to be unambiguously defined regardless of the orientation of this surface. Even if the latter undergoes rotation, twisting and further deformation during the merging, it is therefore safe to take $ I = $ const. when looking for a solution of the variational principle. In other words, even if both $ \int_{\Omega} P_{J} d^3\mbox{x} $ and $ I $ may depend on time, the constraint $ \nabla \cdot \textbf{j} = 0 $ allows us to describe the relaxed state at a given time as a configuration which minimizes $ \int_{\Omega} P_{J} d^3\mbox{x} $ with the constraint of given $ I $.

As for $ \int_{\Omega} P_{J} d^3\mbox{x} $, we may write: $ \int_{\Omega} P_{J} d^3\mbox{x} = \int_{\Omega} \eta_{//}^{-1} \vert \textbf{E} \vert^2 d^3\mbox{x} =  \eta_{//}^{-1} \vert \textbf{E} \vert^2 V_{\Omega} $, where $ \vert \textbf{E} \vert = M_{A} \vert \textbf{B} \vert c_{A} $ and $ V_{\Omega} = \left( R L_{ext} \right) \cdot R $ is the volume of integration. (Having in mind spheromak-like plasmoids with negligible $ \textbf{j} \wedge \textbf{B} $, we have neglected the contribution of $ \eta_{\perp} $. Moreover, nothing essential changes in the following if we replace the length $ R $ which multiplies the area $ R L_{ext} $ in the expression for $ V_{\Omega} $ with some other length, depending on the detailed reconnection model. Finally, here it is not required that $ \eta_{//} $ follows Spitzer's law). Moreover, $ R L_{ext} = \frac{\Delta \Phi}{\vert \textbf{B} \vert} $ where \eqref{FARADAY} implies $ \Delta \Phi = V \cdot \Delta t $ with $ \Delta t = \frac{L}{v_{C}} $ and $ \frac{v_{C}}{c_{A}} = M_{A} \frac{L}{R} $. It follows that $ \int_{\Omega} P_{J} d^3\mbox{x} \approx V \cdot I $, where $ I = M_{A} \eta_{//}^{-1} \vert \textbf{B} \vert c_{A} R^2 $.  

Remarkably, minimization of $ V \cdot I = \min. $ with constant $ I $ is equivalent to minimization of $ V = \min. $ with constant $ I $. According to the reciprocity principle for isoperimetric problems \cite{Elsgolts}, this variational problem has the same solution of the variational problem $ I = \max. $ with constant $ V $. In turn, the latter problem has the same solutions of $ V \cdot I = \max. $ with constant $ V $, i.e. $ \int_{\Omega} P_{J} d^3\mbox{x} = \max. $ with constant $ V $. Physically, when a given amount of electric current $ I $ flows across a conductor we expect the electrons to follow the path which minimizes the resistance $ R_{\Omega} $, so that the Joule dissipated power $ \int_{\Omega} P_{J} d^3\mbox{x} = R_{\Omega} I^2 $ gets also minimized: this is a simple interpretation of Kirchhoff's principle. At fixed voltage $ V $, in contrast, minimization of $ R_{\Omega} $ corresponds to maximization of $ \int_{\Omega} P_{J} d^3\mbox{x} = \frac{V^2}{R_{\Omega}} $. As the merging goes on, electrons involved in Joule dissipation look for the path of lesser resistance. We justify the constraint $ V = $ const. below. 

Substitution of $ I = M_{A} \eta_{//}^{-1} \vert \textbf{B} \vert c_{A} R^2 $ in the variational problem $ I = \max. $ with constant $ V $ implies 

\begin{equation}
\label{hlp1}
M_{A} \eta_{//}^{-1} \vert \textbf{B} \vert c_{A} R^2 = \max \quad \mbox{for fixed} \quad V
\end{equation}

The definitions of $ E_{M} $ and $ M $ give $ E_{M} = \frac{\vert \textbf{B} \vert ^2}{2 \mu_{0}} V_{\Omega} $ and $ M = n_{i} m_{i} V_{\Omega} $ respectively, then $ c_{A} = \sqrt{\frac{2E_{M}}{M}} $. Moreover, the definitions of $ V $ and $ \Delta t $, together with \eqref{FARADAY}, give $ V = L_{ext} \vert \textbf{E} \vert = L_{ext} \cdot \frac{\vert \textbf{B} \vert R}{\Delta t} = \frac{L_{ext}}{L} \cdot v_{C} \vert \textbf{B} \vert R $. For spheromak-like \cite{Bellan}, approximately spherical plasmoids we take $ L_{ext} = L $. (It is reasonable to make the same assumption even if the system is in a non-Taylor DB state). After division of both sides of \eqref{hlp1} by $ \vert \textbf{B} \vert R $, \eqref{Mcons} and \eqref{hlp1} lead therefore to:

\begin{equation}
\label{hlp2}
M_{A} \eta_{//}^{-1} \sqrt{E_{M}} R = \max \quad \mbox{for fixed} \quad v_{C}
\end{equation}

Let us justify the constraint $ V = $ const. invoked above. We recall that $ R \leq L $. Then, maximization of $ M_{A} \eta_{//}^{-1} \sqrt{E_{M}} R $ in \eqref{hlp2} implies that we replace $ R $ with $ L $, or, equivalently, $ v_{C} $ with $ v_{in} $. The applied voltage per unit length in the micro-scale region $ v_{in} \vert \textbf{B} \vert $ reduces therefore to $ v_{C} \vert \textbf{B} \vert $ where both $ v_{C} $ and $ \vert \textbf{B} \vert $ are separately assigned outside the region. 

Now, it comes to $ M_{A} $. After division of both sides by $ R = L $, \eqref{hlp2} reduces to $ M_{A} \eta_{//}^{-1} \sqrt{E_{M}} = \max . $ with constant $ \frac{1}{\Delta t} $, i.e. with constant $ \Delta t $. Moreover, Joule dissipation decreases $ E_{M} $. (This agrees with \eqref{rec01}: the stronger the dissipation, the lower $ E_{M} $, the lower $ c_{A} $ at constant $ M $, the larger the L.H.S. of \eqref{rec01} at given $ v_{C} $, the more likely the merging). Then, in its relaxation the system has to pursue a further, constrained maximization of $ M_{A} \eta_{//}^{-1} $. The latter implies separate maximization of $ \eta_{//}^{-1} $ (electrons look for the path of lesser resistance) and of $ M_{A} $, i.e.:

\begin{equation}
\label{hlp3}
M_{A} = \max \quad \mbox{for fixed} \quad \Delta t
\end{equation}

At a first glance, \eqref{hlp3} seems rather puzzling. We have described the evolution of the system in time as a succession of relaxed states, each satisfying Kirchhoff's principle: accordingly, both $ \int_{\Omega} P_{J} d^3\mbox{x} $ and $ I $ depend on time. In constrast, the duration $ \Delta t $ of the interaction between plasmoids is obviously a property of the entire evolution of the system, from the very beginning when two plasmoids are still far away from each other to the end, when only one structure survives.  

The conundrum is solved by the results of Ref. \cite{Liu0} (where $ M_{A} $ is dubbed 'local reconnection rate'): $ M_{A} $ is a complicated function of the geometry of the magnetic field lines, but its maximum is a quite broad one - i.e., near the maximum $ M_{A} $ depends very weakly on the detailed geometry of field lines. Of course, the motion of the merging plasmoids modifies the geometry of the field lines. As a consequence, once $ M_{A} $ has achieved a maximum it sticks pretty much to the same value as the merging goes on. Moreover, the value computed in \cite{Liu0} depends on geometry only, i.e. it does not depend on $ \Delta t $; i.e., different plasmoid-plasmoid merging events with different duration $ \Delta t $ correspond the same maximum value of $ M_{A} $. Thus, \eqref{hlp3} leads to \eqref{rec02}. 

The model of \cite{Liu0} lacks the reason why $ M_{A} $ attains a maximum. This is likely due to the fact that this model does not take into account the irreversible conversion of upstream energy into heat. Our \eqref{variational} fills the gap, as it implies that electrons which flow across an electric conductor (like e.g. our system) and are subject to a given voltage per unit length (like $ v_{in} \vert \textbf{B} \vert $) tend to follow the path of lesser resistance and therefore to maximize Joule dissipated power (hence $ M_{A} $), just like they do in the lab. This is the large-scale constraint hinted at above. Remarkably, this constraint holds regardless of detailed ion dynamics, and is therefore useful when it comes to the description of merging plasmoids.

The result has a further counterpart in 3D problems where $ \textbf{B} = 0 $ nowhere and $ E_{//} \equiv \frac{\textbf{E} \cdot \textbf{B}}{\vert \textbf{B} \vert^2} \neq 0$. In spheromak-like plasmas we write $ \int_{\Omega} P_{J} d^3\mbox{x} = \int_{\Omega} \textbf{E} \cdot \textbf{j} d^3\mbox{x} \propto \int_{\Omega} \textbf{E} \cdot \textbf{B} d^3\mbox{x} \approx \left( \Delta \Phi \right) \cdot \left( \int E_{//} ds \right) $ where $ \int E_{//} ds $ is computed on a magnetic field line and its maximum value (on the field lines) is $ \propto \frac{\Delta \Phi}{\Delta t} \propto \vert \textbf{E} \vert \propto M_{A}$ \cite{Hesse}; maximization of $ \int_{\Omega} P_{J} d^3\mbox{x} $ reduces again to maximization of $ M_{A} $.

Moreover, the model of \cite{Liu0} provides no explicit dependence of $ M_{A} $ on $ S $. In contrast, Sweet and Parker's model gives $ M_{A} = S^{-\frac{1}{2}} $, and Petschek's model predicts that the maximum value of $ M_{A} $ is equal to $ \frac{\pi}{8 \ln S} $. However, both Sweet and Parker's and Petschek's models are steady state models; but 
in a reconnection layer dominated by the presence of plasmoids, the complexity of the dynamics gives rise to a strongly time-dependent process \cite{Comisso}. 
If $ S $ is small enough, then the maximum value of $ M_{A} $ predicted by these models exceeds the constant value of $ M_{A} $ in \eqref{rec02}. Given the search for a maximum of $ M_{A} $, \eqref{rec02} holds at large $ S $ only, in qualitative agreement with Sec. 5 of Ref. \cite{Cassak}. 

Finally, according to \cite{Liu0} the breadth of the maximum of $ M_{A} $ allows \eqref{rec02} to apply to a wide range of disparate (even relativistic) physical systems with basically the same numerical value of the constant. Indeed, the value of the constant in \eqref{rec02} is 0.2 \cite{Liu0}, i.e. not too far from the value 0.1 widely observed in many problems of reconnection \cite{Cassak}. (Remarkably, the model of \cite{Liu0} assumes $ \beta \ll 1 $ and ensures that the outflow velocity reduces to the Alfvèn velocity for thin reconnection layers. If these assumptions are violated, as e.g. in compressible plasmas where shock waves are present \cite{Mishra}, then different values of $ M_{A} $ are expected \cite{Comisso}). The qualitative nature of our arguments may justify the discrepancy. 

\end{document}